\documentclass{aa}
\usepackage{graphicx}
\usepackage{amsmath,amssymb}
\usepackage{xcolor}
\usepackage{epstopdf}
\usepackage{natbib}
\usepackage{hyperref}
\hypersetup{
    colorlinks=true,
    linkcolor=blue,
    citecolor=blue,
    urlcolor=cyan,
    }
\usepackage{txfonts}
\usepackage{url}

\begin{document}

\title{Can meridional flow variations explain the observed rising/declining phase asymmetry in the solar cycle?}

\author{Soumitra Hazra \inst{\ref{cea}}, Allan Sacha Brun\inst{\ref{cea}}, Laurene Jouve \inst{\ref{irap1} }}

\institute{D\'epartement d'Astrophysique/AIM CEA/IRFU, CNRS/INSU, Univ. Paris-Saclay \& Univ. de Paris Cit\'e 
91191 Gif-sur-Yvette, France \\
email:soumitra.hazra@cea.fr, soumitra.hazra@gmail.com\label{cea}\\
email:sacha.brun@cea.fr\\
\and Toulouse Univ., Institut de Recherche en Astrophysique et Planétologie, CNRS, CNES, Toulouse, France \label{irap1}\\
email:ljouve@irap.omp.eu\\}

   \date{Received; accepted }

\abstract{Accurate forecasting of the 11-year solar cycle remains a central challenge in solar physics, with major implications for space weather prediction. A striking feature of the cycle is its asymmetry between the rising and declining phases, with the decay phase typically lasting much longer. This asymmetry could be due to variations in the Sun’s meridional circulation, though whether these variations are primarily deterministic—driven by Lorentz-force feedback—or stochastic remains debated.} {We aim to determine whether deterministic variations, stochastic fluctuations, or a combination of both in the meridional circulation can reproduce the observed rise–decay asymmetry of the solar cycle.}{We performed kinematic flux-transport dynamo simulations incorporating three classes of time-dependent meridional flow profiles: (i) deterministic variations, (ii) stochastic fluctuations, and (iii) hybrid combinations. To evaluate cycle asymmetry, we analyzed four diagnostics: the rise-to-decay time ratio, and correlations of cycle amplitude with rise time, rise rate, and decay rate near the preceding minimum.}{Solar cycle asymmetry is highly sensitive to the temporal evolution of the meridional flow. When both the meridional flow and the Babcock–Leighton mechanism are stochastic, the model fails to produce cycles with decay times consistently longer than rise times. Physically motivated deterministic variations, inspired by Lorentz-force feedback and interpreted as a response to emergence and equatorward migration of active regions (i.e., the butterfly diagram), are able to reproduce the observed asymmetry. A representative case is obtained when the flow is modulated as $\delta v \sin^2(2\theta_{\mathrm{max}})$, where $\delta v$ is the modulation amplitude and $\theta_{\mathrm{max}}$ the latitude of the toroidal field maximum. This formulation captures the essential feedback effect in the model: the meridional flow weakens near cycle maximum, remains suppressed afterward, and subsequently recovers.  Hybrid scenarios, combining deterministic and stochastic variability along with Babcock–Leighton fluctuations, are also able to  reproduce rise–decay asymmetry. Across all cases, a robust positive correlation emerges between cycle amplitude and rise rate, while correlations with rise time and decay rate remain weak but significant.}
{Meridional circulation variability plays a critical role in shaping solar cycle asymmetry in flux transport dynamo model scenario. Improved observational constraints on its spatiotemporal behavior are essential. Incorporating such variability into forecasting tools—such as Solar Predict—can enhance their physical realism and predictive skill.}

\titlerunning{asymmetry and meridional flow}
      
\maketitle 

\section{Introduction}
The 11-year solar magnetic activity cycle is a prominent manifestation of magnetohydrodynamic processes operating within the Sun's convection zone \citep{Parker1955}. A defining feature of this cycle is the periodic emergence and disappearance of sunspots, accompanied by systematic reversals of the Sun’s global magnetic field \citep{Charbonneau2020}. This magnetic activity manifests on the solar surface as sunspots, active regions, and large-scale polarity reversals \citep{Hathaway2015}, while in the solar atmosphere, it appears as dynamic phenomena such as solar flares, coronal mass ejections (CMEs), and coronal loops—including their implosive behavior linked to magnetic energy release \citep{Hudson1991, Sarkar2017, Ghosh2017, Brun2017, Samanta2019}. These surface and atmospheric manifestations generate large-scale space weather disturbances that can affect critical technological systems, making it essential to understand the mechanisms governing solar cycle variability \citep{Schrijver2015, Mazumder2018, Spal2022, Roy2023}.

Despite its general periodicity, the solar cycle exhibits considerable variability in amplitude and shape \citep{Gnevyshev1948,Komitov2001, Dash2023, Ghosh2024}. One of the most striking features of the solar cycle is the pronounced asymmetry between its rising and declining phases. Typically, the rising phase—marked by the rapid emergence and growth of sunspots—is shorter and steeper than the more gradual decaying phase. This asymmetry is particularly pronounced in cycles with larger amplitudes \citep{Waldmeier1935}. The Waldmeier effect refers to the empirical observation that stronger solar cycles tend to rise faster and reach their peak activity more quickly than weaker ones, which rise more slowly and decay more gently \citep{Waldmeier1935}. However, this correlation is not consistently present across all solar activity proxies; for instance, \cite{Dikpati2008} reported only a weak correlation in sunspot area dataset. Conversely, a strong and consistent correlation has been found between the rise rate and the amplitude of the cycle \citep{Cameron2008}. Extending beyond the Sun, \cite{Garg2019} demonstrated that the Waldmeier effect may also hold for other solar-like stars, suggesting it may be a general feature of stellar magnetic activity cycles. Understanding the physical origin of this asymmetry remains a fundamental challenge in solar dynamo theory and a key ingredient for improving solar cycle predictions.

Flux-transport dynamo models, particularly those based on the Babcock–Leighton (BL) mechanism, provide a promising framework for understanding the solar cycle \citep{Dikpati1999, Nandy2002, Jouve2008, Bhowmik2018, Brun2022}. In the mean-field flux transport dynamo scenario, the solar cycle is sustained through the continuous transformation between toroidal and poloidal magnetic field components \citep{Parker1955a}. This cyclic regeneration is governed by two fundamental processes: the shearing of poloidal field lines by differential rotation, known as the $\Omega$-effect \citep{Parker1955a}, which generates the toroidal field, and the generation of poloidal field from the decay of tilted bipolar active regions at the solar surface, a process encapsulated in the Babcock–Leighton (BL) mechanism \citep{Babcock1961, Leighton1969}. Differential rotation and meridional circulation regulate the conversion and transport of magnetic flux. Further insights into solar and stellar dynamo mechanisms can be found in a range of papers and reviews \citep{Charbonneau2020, Brun2015, Strugarek2017, Panja2021, Noraz2022}.  While successful in explaining several features of the solar cycle, Babcock–Leighton dynamo models often struggle to reproduce the observed cycle asymmetry, raising questions about the role of time-dependent processes.

Two key processes that shape solar cycle variability and asymmetry are fluctuations in the Babcock–Leighton mechanism and changes in meridional circulation. The BL mechanism is inherently stochastic due to scatter in sunspot emergence latitudes and tilt angles. These fluctuations contribute to variations in cycle amplitude and timing, and are closely linked to features such as the Waldmeier effect \citep{Karak2011, Sanchez2014, Pal2023, Tripathi2021, Saha2025}.  Meanwhile, the large-scale meridional flow plays a crucial role in setting the cycle's strength, shape, and duration in flux transport dynamo models \citep{Jouve2007, Jouve2011, Lopes2009, GHazra2023, Vashishth2024}. Please note that, \cite{Zhang2022} recently developed a new distributed dynamo model in which toroidal field generation occurs throughout the convection zone, where meridional circulation plays a very minor role in controlling cycle periodicity and amplitude \citep{Jiang2025}. Nevertheless, the intrinsic nature of the solar meridional circulation—whether it is predominantly deterministic or significantly influenced by stochastic processes—remains a subject of active debate in solar dynamo theory. Surface observations and helioseismic measurements have revealed the presence of a large-scale, poleward meridional flow at the Sun’s surface, with evidence suggesting cycle-dependent variations in its amplitude and structure \citep{Ulrich2010, Basu2010, Hathaway2010, Komm2015, Mahajan2021}. These systematic changes point to a deterministic component, likely driven by dynamo-related magnetic feedback mechanisms. Specifically, as the toroidal magnetic field strengthens during the solar cycle, the associated Lorentz force—especially the magnetic tension component—can modify meridional flow patterns through back-reaction on the plasma \citep{Rempel2006, Lopes2009, GHazra2017}. However, short-term fluctuations, hemispheric asymmetries, and discrepancies between different observational techniques also suggest a stochastic influence, possibly stemming from turbulent convection and localized magnetic activity \citep{Svanda2007, Gizon2020}. This uncertainty—whether the flow is primarily deterministic, stochastic, or a hybrid—poses a key challenge for solar dynamo modeling. Furthermore, polar field imbalance and asymmetric meridional flow at cycle minima have been suggested as precursors to hemispheric asymmetries in subsequent cycles \citep{Bhowmik2019, Lekshmi2019}. Thus, understanding the coupled and possibly nonlinear dynamics of the BL mechanism and meridional circulation—both in their deterministic and stochastic manifestations—is essential for accurately modeling the solar magnetic cycle and explaining its observed amplitude modulation and asymmetric evolution \citep{Brun2004, Brun2011, Hung2015, Nandy2023, Jouve2025}.

Previous studies have typically explored either deterministic feedback or stochastic fluctuations in the meridional flow independently, without examining their combined effects. Similarly, while the stochastic nature of the Babcock–Leighton mechanism is recognized as a source of solar cycle variability, its role in shaping cycle asymmetry when coupled with time-dependent flow remains underexplored. In this study, we address this gap by systematically investigating the interplay between deterministic and stochastic processes in both the meridional circulation and the BL mechanism using a kinematic flux-transport dynamo model. We incorporate a prescribed slowdown of the meridional flow at active latitudes to mimic the Lorentz-force feedback on large-scale flows, and introduce random fluctuations in both the flow and the BL poloidal source. Within this kinematic framework, the modulation of the flow is implemented deterministically, without solving the full dynamical momentum equation. Our focus is on global (not hemispheric) cycle asymmetry, quantified through four metrics: rise–decay time difference and the relationships of cycle amplitude with rise time, rise rate, and decay rate near the preceding minimum. Using both solar cycle observations and dynamo simulations, we explore how variability in the meridional flow and BL mechanism influence solar cycle amplitude, timing, and asymmetry—shedding light on features such as the Waldmeier effect and broader magnetic activity variations.  We provide details about our observational analysis of solar cycle asymmetry in Sect. 2 followed by analysis of asymmetry obtained from our kinematic flux transport solar dynamo model in Sect. 3. Finally, in the last section, we present our conclusions.

\section{Observational Analysis of Solar Cycle Asymmetry and the Waldmeier Effect}
\begin{figure*}[!h]
\centering
\begin{tabular}{cc}
\includegraphics*[width=1.0\linewidth]{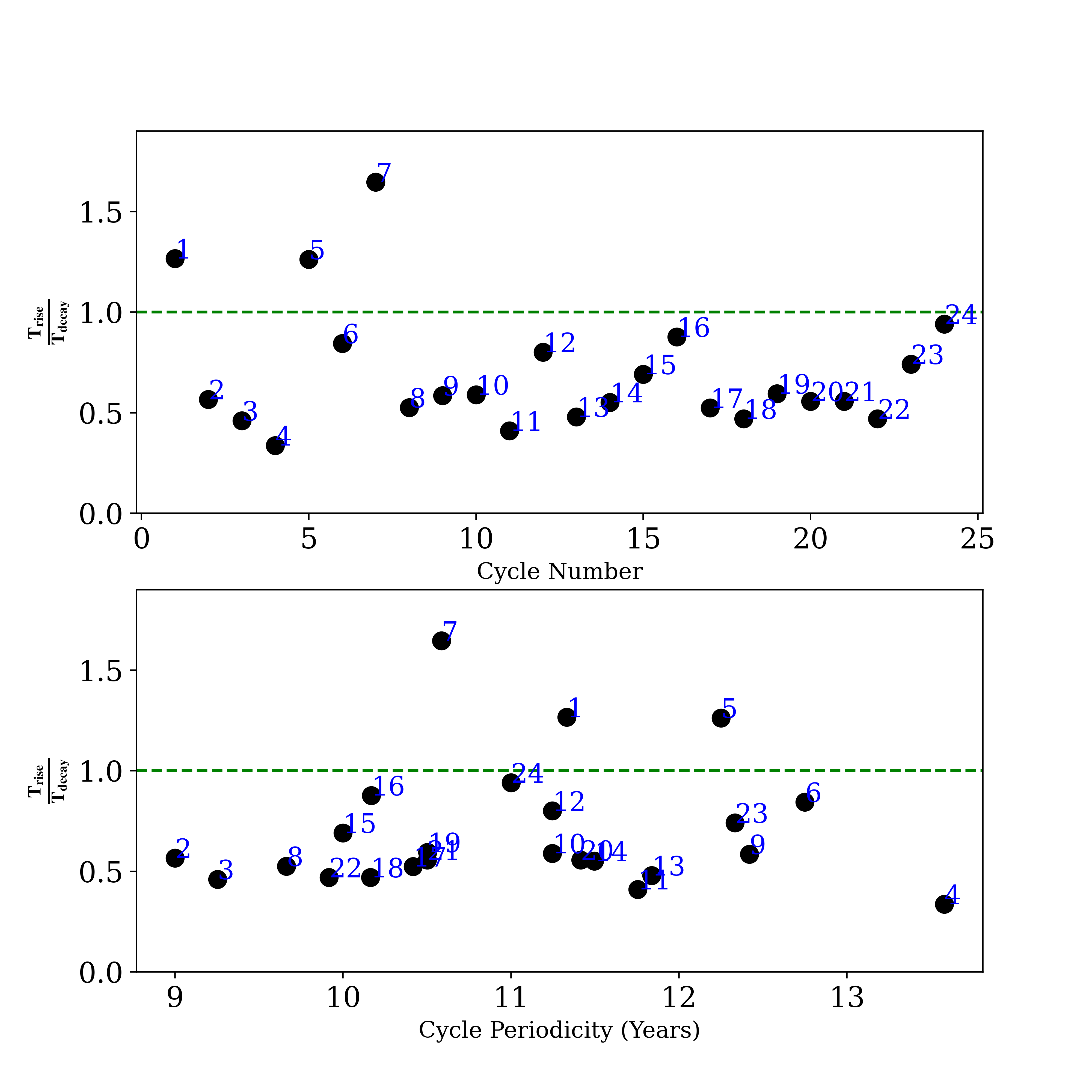}
\end{tabular}                
\caption{\footnotesize{Observational evidence of solar cycle asymmetry in terms of rise and decay times. Top panel: Scatter plots of the ratio $\frac{T_{\mathrm{rise}}}{T_{\mathrm{decay}}}$ versus cycle number, obtained from the sunspot number (SN) data. Bottom panel: Scatter plots of $\frac{T_{\mathrm{rise}}}{T_{\mathrm{decay}}}$ versus cycle periodicity, based on the sunspot number data. The corresponding cycle numbers are annotated in blue in each scatter plot.}}
\label{fig:rise_decay}
\end{figure*}

For our observational analysis of the solar cycle asymmetry, we utilized monthly smoothed sunspot number and sunspot area data. Specifically, we employed the 13-month smoothed sunspot number data obtained from the World Data Center SILSO, and the calibrated sunspot area data compiled by \citet{Mandal2020}. Using these datasets, we computed the rise time and decay time of each solar cycle based on both sunspot number and sunspot area records. In this study, we define the rise time as the duration from solar minimum to the subsequent maximum, and decay time as the duration from solar maximum to the following minimum. Our observations indicate that, for the majority of solar cycles, the decay time exceeds the rise time (see top panel of Figure~\ref{fig:rise_decay}). In the case of the 13-month smoothed sunspot number data, we find that the rise time was longer than the decay time in only a few cycles—namely, cycles 1, 5, and 7.  The ratio of rise to decay times  ($T_{rise}/T_{decay}$)
in the observed sunspot number series exhibits a mean value of $0.70 \pm 0.31$, indicating that on average the rise phase is substantially shorter than the decay phase. The median value, 0.60, further highlights the tendency of most cycles to display pronounced temporal asymmetry, with a small number of anomalous cycles contributing to the larger mean. To account for the influence of these outliers, we recomputed the mean after excluding values lying outside one standard deviation from the initial mean. This filtering removes the most extreme cases that disproportionately inflate the average. The resulting mean of $0.61 \pm 0.15$, together with its reduced spread, provides a more representative estimate of the typical rise-to-decay time ratio across solar cycles. Bottom panels of Figure~\ref{fig:rise_decay} indicate that there is no significant correlation between the $T_{rise}/T_{decay}$ and the cycle period. We find similar kind of behaviour also from the sunspot area data set. In summary, this asymmetry—where the decay time is typically greater than the rise time—emerges as a consistent feature across most solar cycles.

Rise-decay asymmetry of the solar cycle have been extensively analyzed in earlier studies \citep{Hathaway1994, Hathaway2011, Hathaway2015, Jiang2018}. For reference, \cite{Hathaway1994} and \cite{Hathaway2015} provided an analytic fit to all observed sunspot cycles using the expression $F(t) = A \left( \frac{t - t_0}{b} \right)^3 
\left[ \exp\left( \frac{t - t_0}{b} \right)^2 - c \right]^{-1}$, where A denotes the cycle amplitude, $t_0$ is the start time, b is the rise time , and c is an asymmetry parameter. An average cycle is well fit with $A = 195$, $b = 56$, $c = 0.8$, and $t_0 = -4$ months (prior to minimum). This formulation reproduces the asymmetric shape of solar cycles, characterized by a cubic rise and a Gaussian-like decay, leading to a shorter rising phase and a longer declining phase with an average rise-to-decay time ratio of $\approx$ 0.6. Our observational results yield a comparable ratio, confirming the robustness of this asymmetric behavior. Nevertheless, we acknowledge that defining the rise time remains somewhat challenging, as the solar maximum can extend over several years depending on the adopted smoothing and assumed cycle boundary criteria \citep{Jiang2018, Mandal2020}.

\begin{figure*}[!ht]
\centering
\begin{tabular}{cc}
\includegraphics*[width=1.0\linewidth]{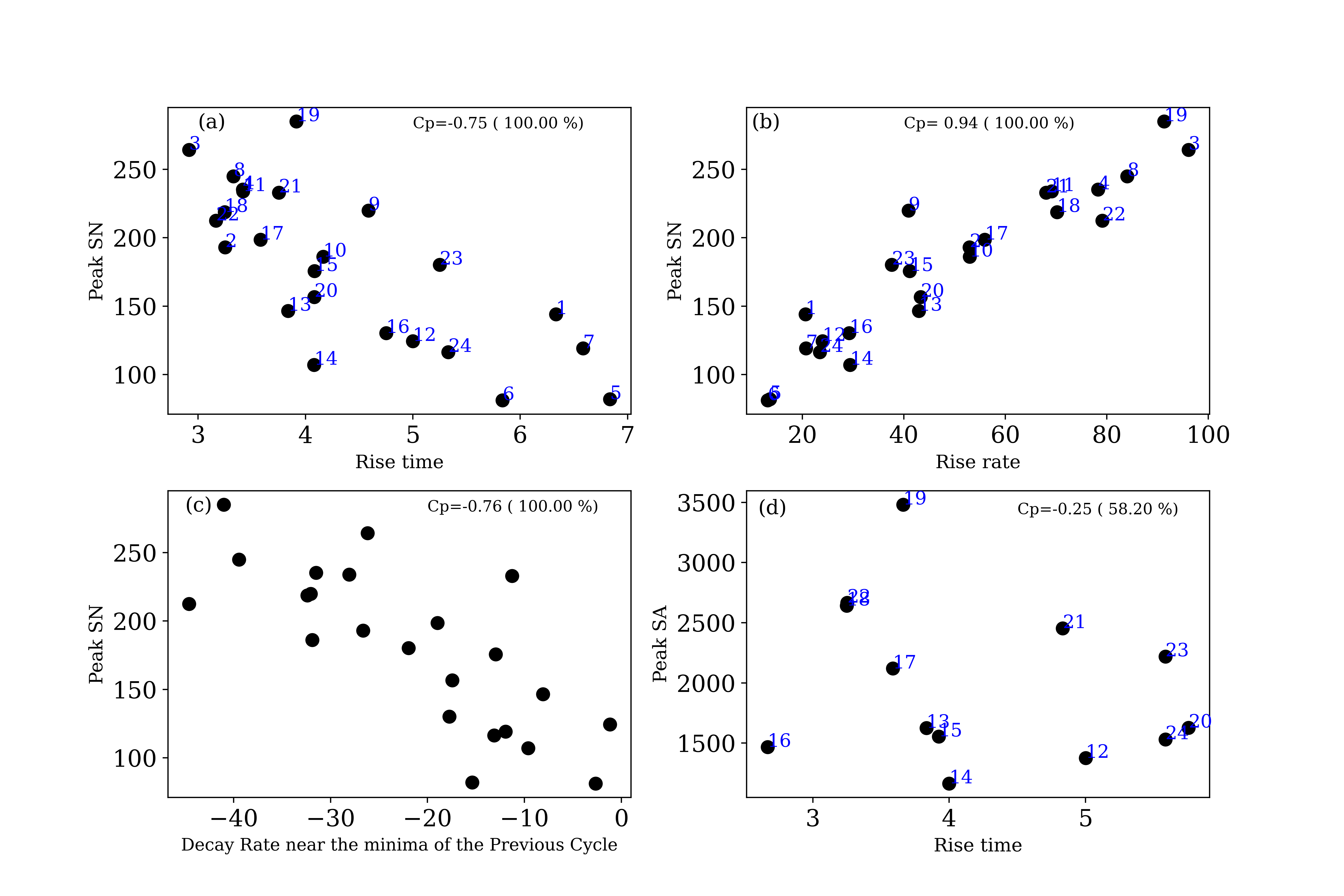}
\end{tabular}                
\caption{\footnotesize{Observational evidence of solar cycle asymmetry. (a) Scatter plots of peak sunspot number versus rise time (in years). (b) Scatter plots of peak sunspot number versus rise rate (in units of sunspot number per year). (c) Scatter plots of peak sunspot number versus decay rate near the end of the previous cycle minimum (in the same units as the rise rate). (d) Scatter plots of peak sunspot area versus rise time (in years). The corresponding cycle numbers are annotated in blue in each scatter plot.}}
\label{fig:wald}
\end{figure*}

Next, we aim to investigate other observed features of the solar cycle, such as the Waldmeier effect, using sunspot number and sunspot area time series data. The Waldmeier effect can be considered a specific manifestation of solar cycle asymmetry. While general asymmetry refers to the tendency of solar cycles to exhibit a shorter rise time and a longer decay time, the Waldmeier effect introduces a quantitative relationship between this asymmetry and the strength of the cycle. In particular, the first form of the Waldmeier effect describes an anticorrelation between the rise time and the amplitude of the cycle—stronger cycles tend to rise more quickly \citep{Waldmeier1935}. The second, more robust form refers to a positive correlation between the rise rate and the cycle amplitude \citep{Cameron2008}. While the first correlation (between rise time and cycle strength) is relatively weak and has proven difficult to establish consistently across all datasets \citep{Dikpati2008, Karak2011}, the second correlation (between rise rate and amplitude) is more reliable and has been consistently observed across various solar proxies. Thus, the Waldmeier effect not only reflects the asymmetric shape of solar cycles but also connects it directly to their intensity, making it a valuable diagnostic feature for understanding solar cycle dynamics.

We have analyzed our monthly sunspot number and calibrated sunspot area datasets to investigate the validity of these correlations. Panel (a) of Figure~\ref{fig:wald} show the relationship between cycle amplitude and rise time obtained from the sunspot number dataset. We find a strong negative correlation (correlation coefficient -0.76) between the peak sunspot number and the rise time when using the sunspot number dataset. Panel (b) of Figure~\ref{fig:wald} illustrate the relationship between cycle amplitude and rise rate obtained from the sunspot number dataset. The rise rate was calculated as the slope during the early rising phase of each cycle. In this case, we observe a strong positive correlation (correlation coefficient 0.94) between cycle amplitude and rise rate in both datasets, reaffirming the conclusions of \citet{Cameron2008}.

Next, we investigate the relationship between the decay rate of the previous cycle and the amplitude of the subsequent cycle. Panel (c) of Figure~\ref{fig:wald} illustrate this relationship obtained from the sunspot number dataset. The decay rate was calculated as the slope between two points separated by two years, with the second point taken one year prior to the cycle minimum. We find a significant negative correlation (correlation coefficient -0.76) between the decay rate during this late phase of the cycle and the amplitude of the following cycle in both datasets. However, when we calculate the decay rate using the entire descending phase of the solar cycle, we do not observe any clear correlation with the next cycle's amplitude. This result suggests that the decay rate near the end of a cycle may serve as a useful precursor for forecasting the amplitude of the subsequent cycle. Similar conclusions have been reported in earlier studies \citep{GHazra2019}, reinforcing the idea that the tail-end decay behavior carries predictive value. Several other studies have also proposed different precursor methods based on solar observations for solar cycle prediction \citep{Schatten2005, Cameron2007, Petrovay2010, Kumar2022, Jaswal2024}.

However, when using the calibrated sunspot area data, the correlation between cycle amplitude and rise time appears significantly weaker (correlation coefficient -0.25). Panel (d) of Figure~\ref{fig:wald} illustrates this relationship based on the sunspot area dataset. Determining the rise time is particularly challenging for some cycles, especially those exhibiting plateau-like maxima or multiple peaks. Notably, \cite{Karak2011} showed that adopting a slightly different definition of rise time leads to a better correlation between peak sunspot area and rise time. Despite the weaker amplitude–rise time relationship, other correlations—such as those between cycle amplitude and rise rate, and between cycle amplitude and decay rate near the preceding minimum—remain robust in the sunspot area dataset, similar to those observed in the sunspot number data. These well-established observational properties of solar cycle asymmetry and the Waldmeier effect provide the physical motivation for our modeling efforts. In the following sections, we investigate the possible origin of this asymmetry within the framework of a Babcock–Leighton dynamo model that includes a time-dependent meridional circulation. This allows us to explore how variations in the large-scale flow, whether deterministic or stochastic, can reproduce the observed asymmetric evolution of the solar cycle.

\section{Asymmetry from Kinematic Solar Dynamo Models}
\subsection{Model}
We solve following evolution equations for the poloidal field ($A$) and toroidal field ($B_\phi$) in axisymmetric spherical polar coordinate system to model the solar dynamo mechanism:
\begin{eqnarray}
     \frac{\partial A}{\partial t} + \frac{1}{s}(\mathbf{v_p}\cdot\nabla)(s A) &=& \eta_p
    \left( \nabla^2 - \frac{1}{s^2}\right)A + \alpha B\,\,\,, \label{eq:1}\\
    \frac{\partial B}{\partial t} + \frac{1}{r}\left[\frac{\partial}{\partial r}(r v_r B)
    + \frac{\partial}{\partial \theta}(v_\theta B)\right] &=& \eta_t
    \left( \nabla^2 - \frac{1}{s^2}\right)B \nonumber \\
    && +\, s ((\nabla \times [A(r,\theta)\hat{\mathbf{e}}_\phi])\cdot\nabla)\Omega \nonumber \\
    && +\, \frac{1}{r}\frac{d\eta_t}{dr}\frac{\partial}{\partial r}(rB)\,\,\,,
    \label{eq:2}
\end{eqnarray}

where, $\Omega$ is the differential rotation, ${\bf v}_p= \bf{v}_r \hat{r} + \bf{v}_\theta \hat{\theta}$ is the meridional flow, $\hat{e}_\phi$ is the azimuthal unit vector, and $s = r\sin(\theta)$. Following \citet{Chatterjee2004}, we assume different magnetic diffusivities for the poloidal and toroidal components of the magnetic field, denoted by $\eta_p$ and $\eta_t$, respectively. This distinction is motivated by the fact that strong magnetic fields can suppress turbulence, thereby reducing the effective magnetic diffusivity—particularly for the toroidal component, which is generally stronger and more confined.

We adopt diffusivity profiles similar to those used by \citet{Chatterjee2004}. The diffusivity profile for the poloidal component is given by:
\begin{equation}
\eta_{p}(r) = \eta_{RZ} + \frac{\eta_{SCZ}}{2}\left[1 + erf \left(\frac{r - r_{BCZ}}
{d_t}\right) \right]
\label{eq:etap}
\end{equation}
where $\eta_{SCZ} = 2.4 \times 10^{12}$cm$^2$ s$^{-1}$, 
$\eta_{RZ} = 2.2 \times 10^8$ cm$^2$ s$^{-1}$, $r_{BCZ} = 0.7 R_\odot$, 
$d_t = 0.05R_{\odot}$.

Diffusivity of the toroidal component is modeled with a more confined profile, given by::
\begin{eqnarray}
\label{eq:etat}
\eta_{t}(r) = \eta_{RZ} + \frac{\eta_{SCZ1}}{2}\left[1 + erf \left(\frac{r - r^{\prime}_{BCZ}}{d_t}\right) \right] \nonumber \\
+ \frac{\eta_{SCZ}}{2}\left[ 1 + erf \left(\frac{r-r_{TCZ}}{d_t}
\right) \right]
\end{eqnarray} 
with $\eta_{SCZ1}=4\times10^{10}$cm$^2$ s$^{-1}$,   $\eta_{SCZ}=2.4\times10^{12}$cm$^2$ s$^{-1}$,
$r^{\prime}_{BCZ} = 0.72R_{\odot}$ and $r_{TCZ}=0.95 R_{\odot}$. This parametrization effectively localizes lower diffusivity around the region where strong toroidal fields are stored, helping to preserve their integrity against turbulent diffusion. Different diffusivity profiles are adopted to model the suppression of turbulent diffusivity by strong magnetic fields. The toroidal field, being strong and concentrated in localized flux tubes, is subject to less diffusion, whereas the weaker, more diffuse poloidal field experiences stronger diffusion \citep{Choudhuri2003}. The choice of a higher diffusivity for the poloidal field throughout the convection zone ensures that our model operates in the diffusion-dominated regime.

We model the Babcock–Leighton mechanism through a near-surface  $\alpha$ profile, parameterized as follows:
\begin{equation}
    \alpha(r,\theta) = \alpha_{BL}\frac{ \cos \theta }{4} \left[1+\textrm{erf}
    \left( \frac{r-r_1}{d_1}\right)\right] \times\ \left[1-\textrm{erf}
    \left( \frac{r-r_2}{d_2}\right)\right]\,\,\,\,,
    \label{c6.eq-alphastandard}
\end{equation}

where $r_1=0.95R_\odot$, $r_2=R_\odot$, $d_1=d_2=0.025R_\odot$ and $\alpha_{BL}$ controls the amplitude of the poloidal field generation. This profile ensures that the $\alpha$-effect is confined to a thin layer near the solar surface and is antisymmetric about the equator, consistent with the expected characteristics of the Babcock–Leighton process. In our model, the effect of magnetic flux tube eruption is incorporated through a buoyancy algorithm. This algorithm periodically scans the base of the solar convection zone to identify regions where the toroidal magnetic field exceeds a specified threshold. When such a region is detected, the algorithm removes a portion (typically half) of the local toroidal flux and deposits it near the surface layers, where the Babcock–Leighton (BL) poloidal source term is active. This mimics the rise of buoyant magnetic flux tubes and their subsequent emergence at the solar surface, which is a key aspect of the solar dynamo process. Recently, \citep{Biswas2022} demonstrated that toroidal flux loss due to flux emergence can explain why the equatorward drift of the central latitude of the activity belts is similar for all solar cycles, leading to comparable decay phases.

For the differential rotation profile, we use an analytic fit to helioseismic rotation data, following \cite{Chatterjee2004, SHazra2020}:
\begin{equation}\label{DRan}
   \begin{array}{cc}
      \Omega(r,\theta) = 2\pi\Omega_{c} + \pi \left( 1 - \operatorname{erf}\left( \frac{r - r_{tc}}{d_{tc}}  \right)\right) 
      \left( \Omega_{e} - \Omega_{c} + ( \Omega_{p} - \Omega_{e} )\Omega_S(\theta) \right) ,\\
      \\
      \Omega_S(\theta) = a\cos^2(\theta) + (1-a)\cos^4(\theta), \\
    \end{array}
\end{equation}
Here, $\Omega_{c}$, $\Omega_{e}$, and $\Omega_{p}$ represent the rotation frequencies of the core, equator, and pole, respectively. The functional form $\Omega_S(\theta)$ captures the latitudinal dependence of surface rotation. The parameter values used are consistent with those adopted in \cite{SHazra2020}.

The meridional circulation profile ($\mathbf{v}_p$) is constructed by specifying a stream function $\psi$, such that
\begin{equation}
\rho \mathbf{v}_p = \nabla \times (\psi \hat{e}_\phi),
\end{equation}
where $\rho = C(R_\odot/r -0.95)^{3/2}$ is the non-dimensional density stratification, and $\hat{e}_\phi$ is the azimuthal unit vector.  The stream function is explicitly prescribed as:
 \begin{eqnarray}
   r \sin \theta \psi(r, \theta)=   
   \psi_0 (r - R_p) \sin \left[ \frac{\pi (r - R_p)}{(R_\odot - R_p)} \right] \{ 1 - e^{- \beta_1 r \theta^{\epsilon}}\} \nonumber \\
  \times~~  \{1 - e^{\beta_2 r (\theta - \pi/2)} \} e^{-((r -r_0)/\Gamma)^2}
,\end{eqnarray}
where $\psi_0$ determines the amplitude of the circulation. The parameters are chosen as follows: $\beta_1 = 1.5 \times 10^{-8}$ m$^{-1}$, $\beta_2 = 1.8 \times 10^{-8}$ m$^{-1}$, $\epsilon = 2.0000001$, $r_0 = (R_\odot - R_b)/4$, $\Gamma = 3.47 \times 10^8$ m. The amplitude of the meridional circulation at mid-latitudes, denoted by $v_0$, is obtained by normalizing the stream function amplitude $\psi_0$ with the constant $C$, i.e., $v_0 = \frac{\psi_0}{C}$. The circulation extends slightly beneath the convection zone, with a penetration depth set by $R_p = 0.65 R_\odot$. This yields a surface poleward flow ranging from 10–31 m s$^{-1}$ and a weak return flow (1 m s$^{-1}$) at the base of the convection zone, consistent with observational constraints \citep{Komm93, Hathaway1996}.

\subsection{Asymmetry due to fluctuations in the poloidal field }
\begin{figure*}[!ht]
\centering
\begin{tabular}{cc}
\includegraphics*[width=1.0\linewidth]{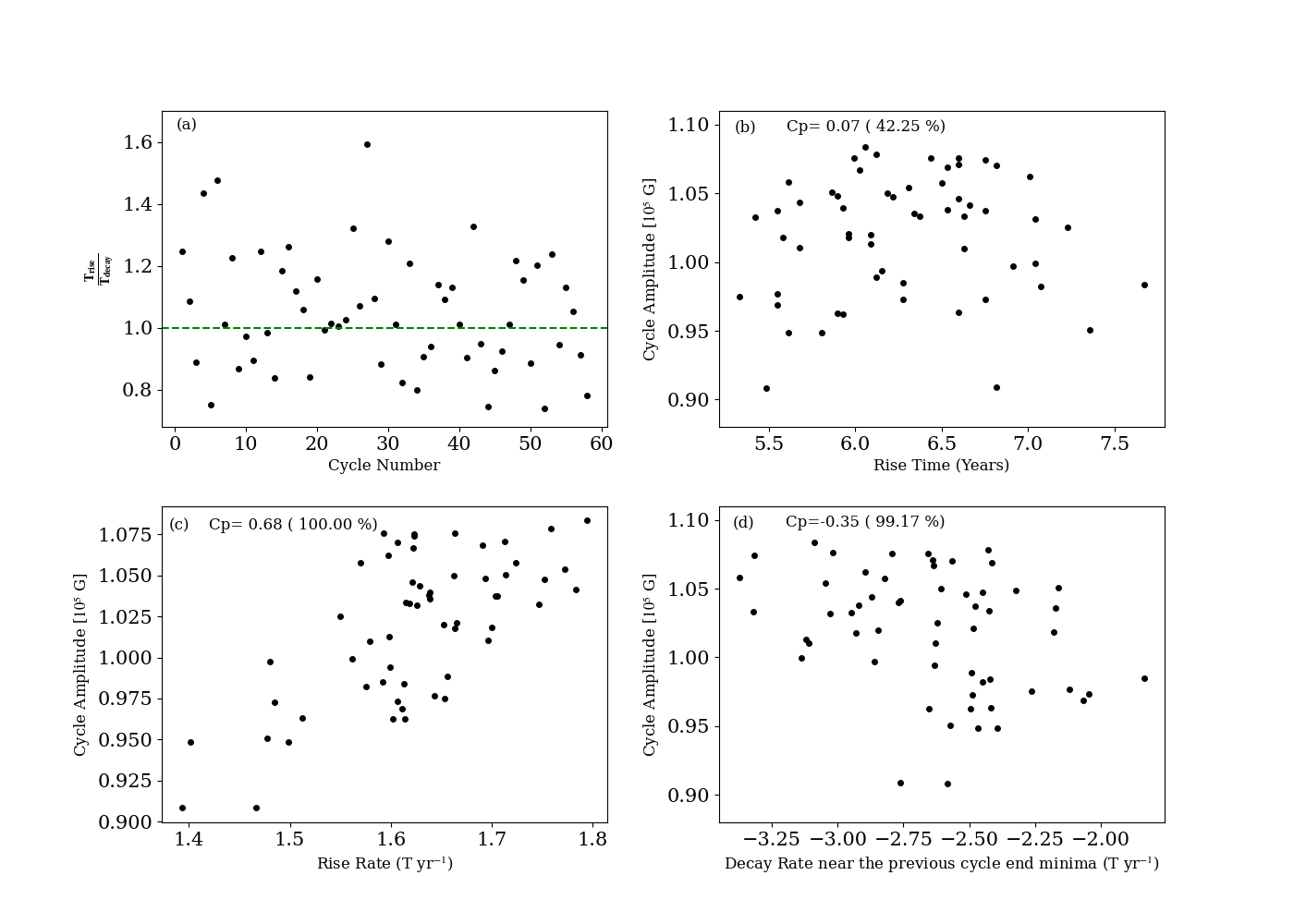}
\end{tabular}                
\caption{\footnotesize{Asymmetries generated from our model by introducing $50 \%$ fluctuations in the poloidal field generation mechanism. (a) Scatter plot of the rise-to-decay time ratio versus cycle number. (b) Scatter plot of cycle amplitude (in Gauss) versus rise time (in years). (c) Scatter plot of cycle amplitude versus rise rate (in units of Tesla/year). (d) Scatter plot of cycle amplitude versus decay rate near the preceding cycle minimum (in Tesla/year).}}
\label{fig:wald-pol}
\end{figure*}
Among the key components of the solar dynamo is the poloidal magnetic field, which acts as the seed for toroidal field generation via differential rotation. The regeneration of the poloidal field—largely governed by the Babcock-Leighton mechanism at the solar surface—is inherently stochastic, as it depends on the emergence and tilt angles of active regions, which are subject to convective buffeting and nonlinear feedback \citep{Jiang2014}. These stochastic variations can lead to hemispheric differences in the strength and structure of the poloidal field, thereby influencing the toroidal field generation in the subsequent cycle. Furthermore, fluctuations or variations in the poloidal source term can also contribute to long-term amplitude modulation of the solar cycle, such as the Maunder Minimum–like episodes \citep{Jiang2007, Passos2014, SHazra2014, SHazra2019, Dash2020}.

First, we investigate the role of poloidal field source fluctuations as a primary driver of asymmetry and amplitude modulation in the solar cycle. To this end, we incorporate stochastic perturbations into the poloidal field source term within a kinematic dynamo model, while keeping the meridional circulation speed fixed. In this study, we introduce a 100 \% fluctuation in the poloidal field generation mechanism, with a correlation time of one year, to emulate the stochastic nature of the Babcock–Leighton process. We mathematically model the Babcock-Leighton mechanism fluctuations as: 
\begin{equation}
\alpha_{BL}= \alpha_{BL}^0[1 + \delta \sigma(t, \tau)]
\end{equation}

where $\alpha_{BL}^0= 50$ m/s, $\delta$ is the percentage level of fluctuation (here around 50 \%), and $\sigma(t, \tau)$ is the time-dependent uniform white noise with an amplitude between +1 and -1 and a coherence time $\tau$. Our adopted value of $\alpha_{BL}^0$ is sufficiently robust to sustain a periodic dynamo solution, irrespective of large fluctuations in the meridional flow strength. Please note that when we refer to a 50 \% fluctuation in the poloidal field generation mechanism, we mean that the value can vary by $\pm 50$ \% around its mean. Our choice of fluctuation amplitude is motivated by observed polar field variability in Wilcox Solar Observatory data, as well as the distribution of eddy velocities in turbulent solar convection simulations \citep{Miesch2008, Racine2011, Passos2012, Noraz2025}. The selected coherence time ($\tau_{cor}$) is consistent with two relevant physical timescales: the rise time of magnetic flux tubes through the turbulent convection zone \citep{Caligari1995} and the redistribution time of active regions by surface flows, both of which are on the order of months to one year.

Figure~\ref{fig:wald-pol}(a) illustrates the variation of the rise-to-decay time ratio, $\frac{T_{rise}}{T_{decay}}$, across solar cycles. Interestingly, we find that the rise time exceeds the decay time in many cycles—a behavior that is rarely observed in actual solar data. This indicates that poloidal field fluctuations alone, in the absence of meridional flow variability, may not fully capture the temporal asymmetry of the solar cycle. Further, Figure~\ref{fig:wald-pol}(b) shows the relationship between cycle amplitude and rise time. Contrary to observations, no clear correlation is found. A moderate positive correlation is observed between the rise rate and the cycle amplitude, as shown in Figure~\ref{fig:wald-pol}(c). Additionally, Figure~\ref{fig:wald-pol}(d) reveals only a weak negative correlation between the amplitude of a cycle and the decay duration near the end of the preceding cycle. Throughout this work, the solar cycle amplitude is represented by the absolute value of the toroidal magnetic field at the base of the convection zone.

In summary, our results indicate that fluctuations in poloidal field generation alone cannot account for the observed asymmetries and amplitude modulations of the solar cycle in the BL flux-transport model. We tested different levels of stochastic variation in the Babcock–Leighton source and confirmed that the results are robust. Additional ingredients—such as variability in meridional circulation or feedback mechanisms—are likely required to reproduce the full range of observed solar cycle characteristics.

\begin{figure*}[!ht]
\centering
\begin{tabular}{cc}
\includegraphics*[width=1.0\linewidth]{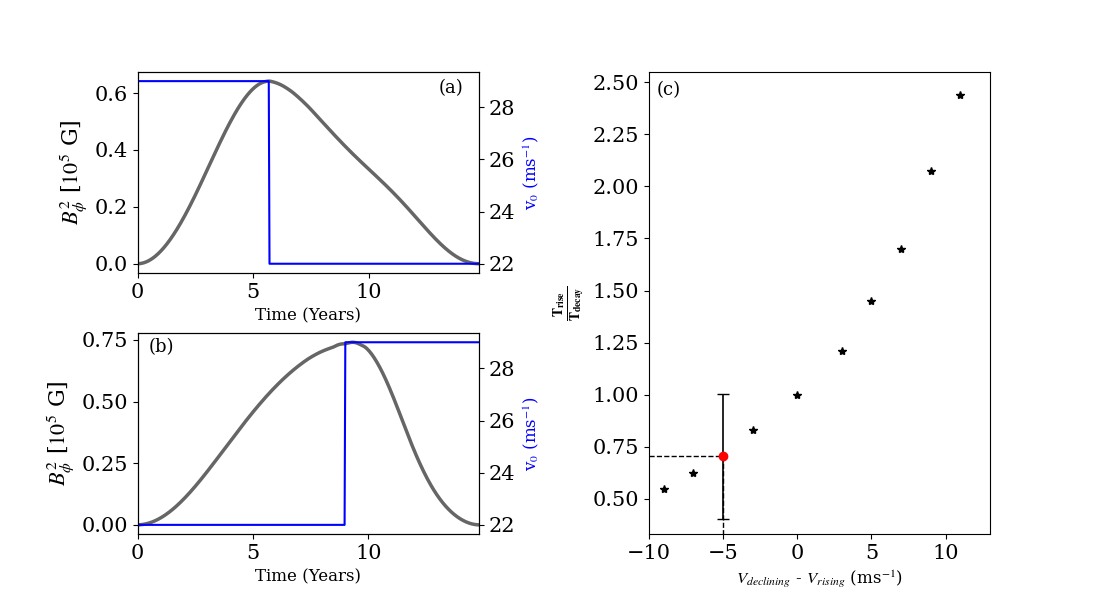}
\end{tabular}                
\caption{\footnotesize{(a) Temporal evolution of $B_\phi^2$ and meridional flow speed when the flow decreases at the cycle maximum. The black curve (left y-axis) shows the variation of $B_\phi^2$ (a proxy for sunspot number), while the blue curve (right y-axis) shows the meridional flow speed decreasing from a higher to a lower value at the cycle peak. (b) Same as (a), but with the meridional flow increasing from a lower to a higher value at the cycle maximum. (c) Variation of the rise-to-decay time ratio as a function of flow asymmetry between the declining and rising phases, expressed as $v_{\rm declining} - v_{\rm rising}$. Red dot corresponds to the mean value of rise-to-decay time ratio.}}
\label{fig:walmerid}
\end{figure*}

\begin{figure*}[!ht]
\centering
\begin{tabular}{cc}
\includegraphics*[width=1.0\linewidth]{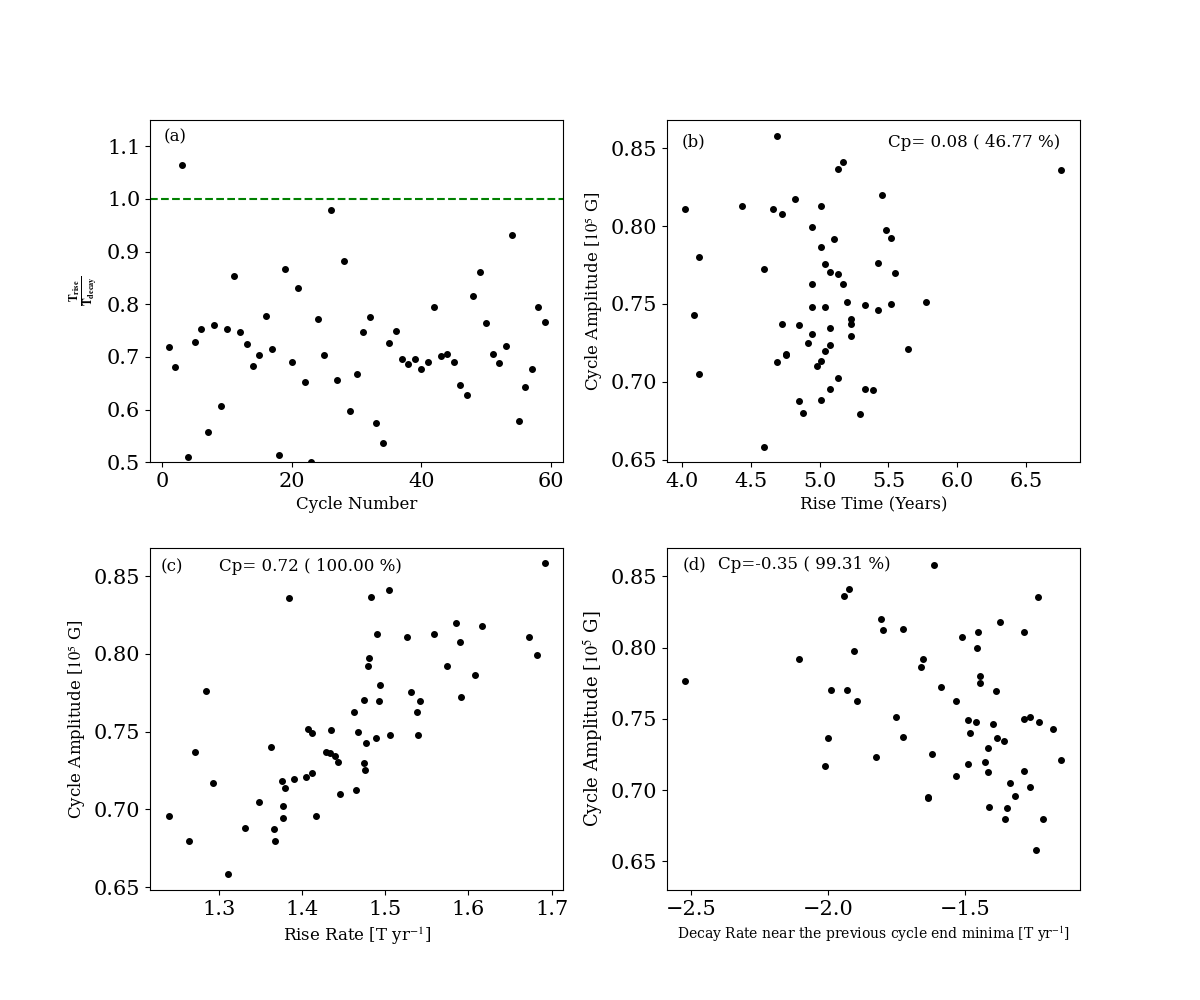}
\end{tabular}                
\caption{\footnotesize{Asymmetry generated from our model when a higher meridional flow speed (29ms$^{-1}$) is applied during the rising phase and a slower flow (23ms$^{-1}$) during the decaying phase. Additionally, $50 \%$ fluctuations are introduced in the poloidal field generation mechanism. The four panels in this figure correspond to the same quantities shown in Figure~\ref{fig:wald-pol}.}}
\label{fig:waldmerid-a1}
\end{figure*}


\begin{figure*}[!ht]
\centering
\begin{tabular}{cc}
\includegraphics*[width=1.0\linewidth]{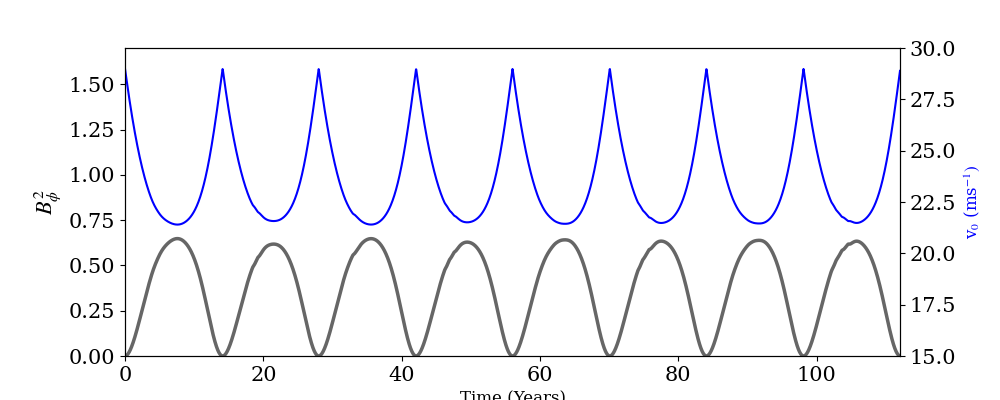}
\end{tabular}                
\caption{\footnotesize{Results when the meridional flow is continuously modulated by Lorentz feedback from the dynamo-generated magnetic field. The black curve (right y-axis) shows the evolution of magnetic energy density, while the blue curve (left y-axis) shows the variation in the amplitude of the meridional flow. }}
\label{fig:lorentz}
\end{figure*}

\begin{figure*}[!ht]
\centering
\begin{tabular}{cc}
\includegraphics*[width=1.0\linewidth]{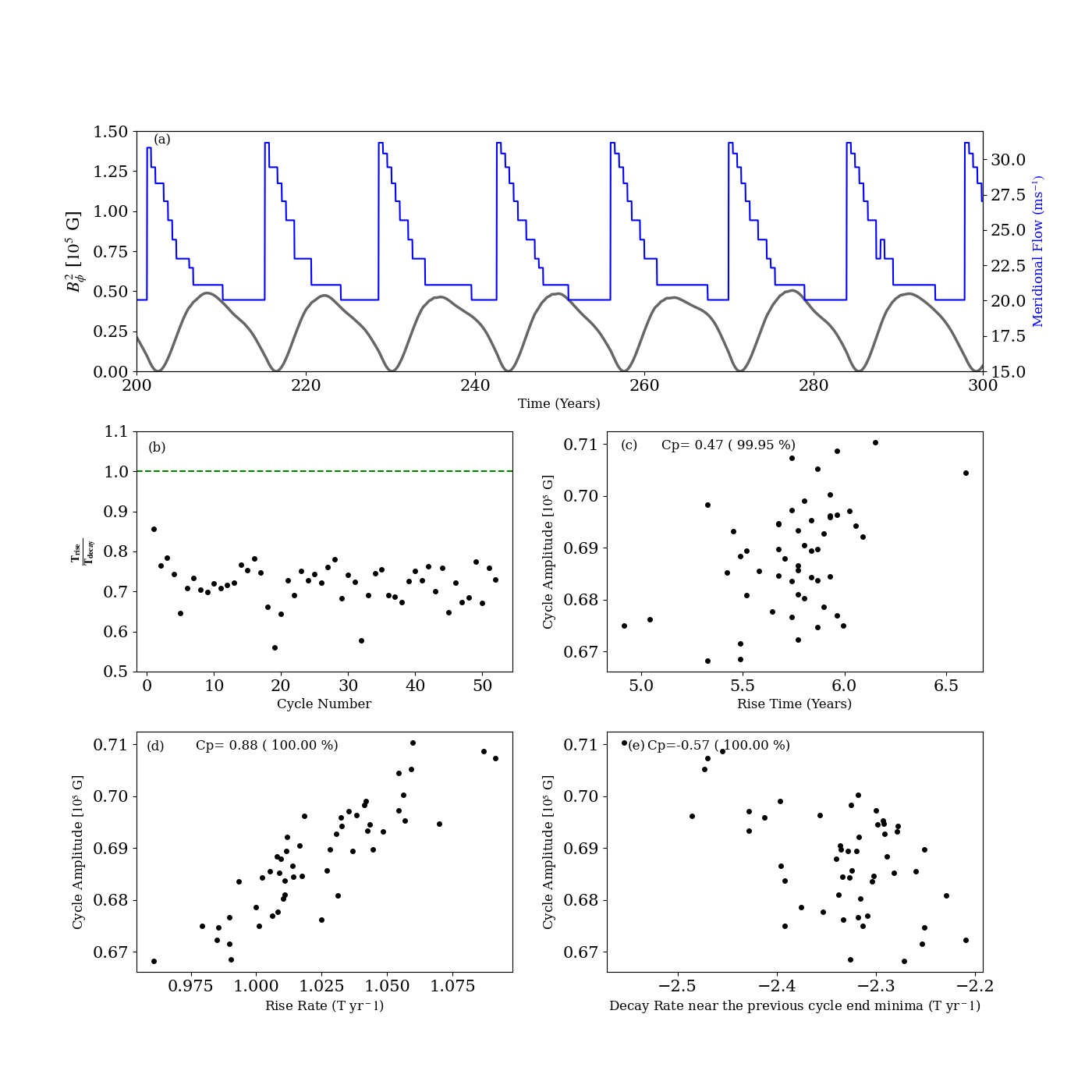}
\end{tabular}                
\caption{\footnotesize{
\textbf{Figure (a), top panel:} Results when the meridional flow is modulated according to the active region eruption belt (butterfly diagram). The modulation is applied using the term $\delta v \sin^2(2\theta_{\mathrm{max}})$, where $\theta_{\mathrm{max}}$ is the most probable eruption latitude—corresponding to the location of the strongest toroidal field. The black curve (right $y$-axis) shows the magnetic energy density evolution, while the blue curve (left $y$-axis) shows the amplitude variation of the meridional flow. \textbf{Figures (b)–(e):} Asymmetry measures derived from the same simulation. These four panels represent the same quantities as in Figure~\ref{fig:wald-pol}. We have also included an additional $50\%$ stochastic fluctuation in the poloidal field generation mechanism. In this scenario, the meridional flow speed weakens near the solar cycle maximum.
}}
\label{fig:butf}
\end{figure*}

\begin{figure*}[!ht]
\centering
\begin{tabular}{cc}
\includegraphics*[width=1.0\linewidth]{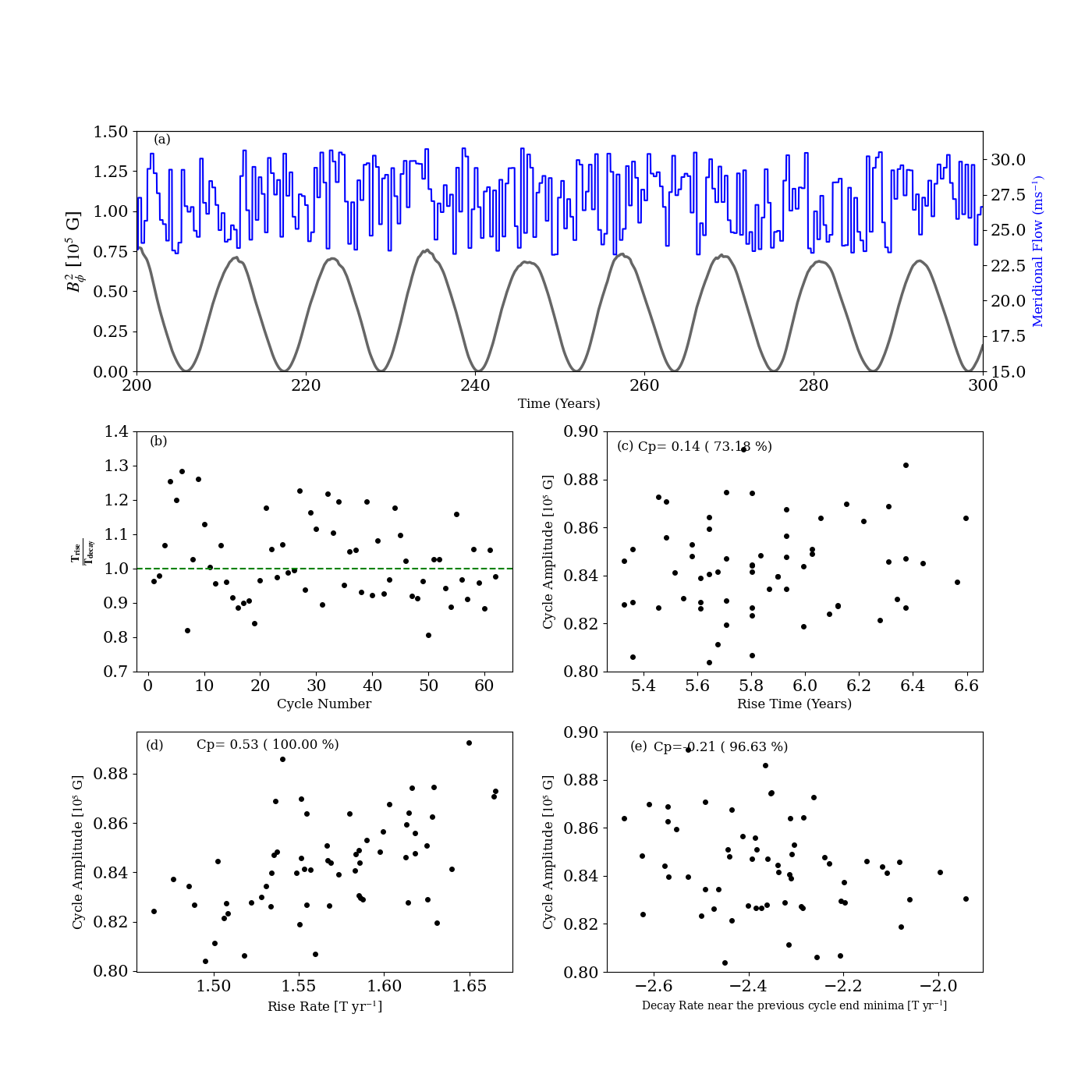}
\end{tabular}                
\caption{\footnotesize{\textbf{Figure (a), top panel:} Results when the meridional flow varies randomly with a coherence time of one month. The black curve (right $y$-axis) shows the magnetic energy density evolution, while the blue curve (left $y$-axis) shows the amplitude variation of the meridional flow. \textbf{Figures (b)–(e):} Asymmetry measures derived from the same simulation. These four panels represent the same quantities as in Figure~\ref{fig:wald-pol}. We have also included an additional $50\%$ stochastic fluctuation in the poloidal field generation mechanism. }}
\label{fig:random-merid}
\end{figure*}
\subsection{Asymmetry due to the deterministic variations in the meridional flow}
Meridional flow plays a key role in determining the duration and amplitude of the solar cycle, with both systematic and cycle-dependent variations influencing dynamo behavior \citep{Choudhuri1995, Jouve2007, Nandy2011, SHazra2016, Hung2017}.  . These variations, often linked to magnetic back-reaction, motivate our analysis of its impact on the model solutions.

To investigate how deterministic variations in meridional circulation affect the asymmetry between the rise and decay phases of the solar cycle, we begin by examining idealized scenarios where the flow speed is different during the rising and declining phases. In the first case, we consider a scenario where the meridional circulation is faster during the rising phase and slower during the declining phase of the cycle. As shown in Figure~\ref{fig:walmerid}(a), this configuration leads to a shorter rise time and a prolonged decay phase. In the second case, we reverse the flow profile—slower in the rising phase and faster in the declining phase. This leads to a longer rise time and a shorter decay time, as illustrated in Figure~\ref{fig:walmerid}(b). To systematically investigate this trend, we compute the ratio of rise time to decay time ($T_{\rm rise}/T_{\rm decay}$) across a range of flow asymmetries, parameterized by the difference in meridional flow speed between the declining and rising phases ($v_{\rm declining} - v_{\rm rising}$). The variation of this ratio is shown in Figure~\ref{fig:walmerid}(c). When the flow difference between the declining and rising phase is approximately $-5$~ms$^{-1}$ in our model, it may reproduce the observed mean rise-to-decay time ratio ($T_{\rm rise}/T_{\rm decay} \approx 0.69$), as indicated by the highlighted point in Figure~\ref{fig:walmerid}(c). This suggests that the temporal asymmetry of the cycle could be captured at this specific flow velocity difference.

 The asymmetry observed in the rise and decay phases of the solar cycle in these scenarios can be directly attributed to the temporal modulation of the meridional flow. When the meridional circulation is faster during the rising phase and slower during the declining phase, the transport of magnetic flux from the equator to the poles is accelerated early in the cycle, leading to a rapid buildup of the poloidal field and, consequently, a swift amplification of the toroidal field through differential rotation. This results in a shorter rise time. However, the slower flow during the declining phase delays the removal and cancellation of residual magnetic flux, thereby extending the decay phase, as seen in Figure~\ref{fig:walmerid}(a). Conversely, when the meridional flow is slower in the rising phase and faster in the declining phase, the initial flux transport is impeded, leading to a delayed buildup of the magnetic fields and a prolonged rise time. The faster flow in the later phase then enhances the flux transport and accelerates the decay, resulting in a shorter decay time, as shown in Figure~\ref{fig:walmerid}(b). These results highlight the critical role of time-dependent meridional circulation in shaping the temporal profile of the solar cycle.

 We now present results for cases in which a fast meridional flow is imposed during the first half of the cycle, followed by a slower flow during the second half, along with 50 \% fluctuations in the poloidal field generation mechanism. In this setup, we fix the meridional flow speed at 29 m/s during the first half of the cycle and at 23 m/s during the second half. Figure~\ref{fig:waldmerid-a1} displays the outcomes from this setup. We clearly observe that the decay time exceeds the rise time for nearly all cycles and in good agreement with the observed values. A strong positive correlation (correlation coefficient 0.72) is found between the cycle amplitude and the rise rate. Additionally, we observe a significant but weak negative correlation (correlation coefficient -0.35) between the cycle amplitude and the decay rate near the preceding cycle minima. However, no notable relationship is found between the cycle amplitude and the rise time.

Next, we consider a more physically motivated, deterministic time-dependent meridional flow profile in which the flow variation arises from the back-reaction of the Lorentz force.  In this formulation, the meridional circulation is modulated in response to the strength of the toroidal magnetic field, with the flow speed decreasing when the toroidal field reaches its maximum. This reduction is a consequence of the magnetic tension exerted by the strong toroidal field, which acts to oppose and suppress the background plasma motion through the Lorentz force. Such a feedback mechanism introduces a self-consistent, nonlinear coupling between the magnetic field and the flow, aligning with both theoretical expectations and helioseismic inferences that suggest weakened meridional circulation near solar maximum. By incorporating this dynamic modulation, we aim to capture more realistic solar cycle behavior and assess the impact of Lorentz-force-driven flow variations on the cycle asymmetry and duration. Mathematically, the modulation is described as: 
\begin{equation}
v_0(t) = v_{\max}\left[1 - \tanh\!\left(\frac{|B_\phi|}{\Delta}\right)\right] 
\end{equation}
where $v_{max}$ is the maximum meridional flow speed (typically near cycle minima), and $\Delta$ controls the degree of flow suppression at cycle maxima. $B_\phi$ value is taken at the base of the convection zone and at a particular latitude of $15^0$.

Figure~\ref{fig:lorentz} presents the results from our model where the meridional flow is continuously modulated by the Lorentz force. In this case, we do not observe any significant asymmetry between the rise and decay phases of the solar cycle. This symmetry arises because the variation in the meridional flow is smooth and nearly symmetric around the cycle maximum, responding directly to the strength of the toroidal magnetic field. As a result, the temporal evolution of the magnetic field remains largely symmetric, in contrast to cases with imposed asymmetric flow profiles. In summary, while these simplified models incorporate key nonlinear feedback mechanisms via the Lorentz force, they are not successful in explaining the observed asymmetry of the solar cycle.

We next consider a deterministic, time- and latitude-dependent meridional flow profile that evolves alongside the sunspot emergence latitudes, effectively mirroring the butterfly diagram. In this formulation, the flow speed is reduced in magnetically active regions—typically around the latitudes where the toroidal field peaks at a given phase of the cycle. This suppression is motivated by two key physical mechanisms: active region inflows, which are converging flows toward sunspot latitudes driven by radiative cooling and pressure deficits in plage regions, and Lorentz force feedback, where magnetic tension and pressure gradients locally inhibit plasma motion. 
As the sunspot belts migrate equatorward throughout the cycle, these mechanisms collectively generate a spatiotemporal modulation of the meridional circulation.

Observational evidence also supports this connection between magnetic activity and flow variability. Previous helioseismic studies have reported solar-cycle-dependent variations in the meridional flow \citep{Basu2010, Zhao2013, Komm2015}. However, \cite{Cameron2010} demonstrated that such variations can be reproduced by superposing inflows toward the activity belts onto an otherwise steady background flow, thereby establishing a direct link to the butterfly diagram. More recently, \cite{Mahajan2023} found that removing active-region inflows reveals only a remaining weak cycle dependency, suggesting that most of the apparent flow variation arises as a response to emergence and equatorward migration of active regions — hence closely linked to the butterfly diagram.

This approach is further supported by helioseismic observations showing flow perturbations beneath active regions, consistent with the idea that enhanced magnetic tension in magnetically active latitudes can locally impede the flow. To implement this, we first identify the latitude between 5$^{\circ}$ and 45$^{\circ}$ where the toroidal field attains its maximum, and define this as the latitude of maximum emergence. The meridional flow is then modified by adding a modulation term, expressed as: 
\begin{equation}
    v_{0} (t) = v_{const} + \delta v \, \sin^{2}(2\theta_{\mathrm{max}}),
\end{equation}
where $v_{const}$ is the constant meridional flow speed, $\delta v$ denotes the amplitude of the modulation and $\theta_{\mathrm{max}}$ represents the latitude where the toroidal field is maximum. The blue curve in Figure~\ref{fig:butf}(a) illustrates the resulting flow structure and its temporal evolution. Please note that here the meridional flow is modulated without directly involving the toroidal field intensity but is tied to the location of maximum sunspot activity. This allows us to incorporate flow suppression near eruption belts and capture realistic temporal evolution of the flow. As expected, the meridional flow speed decreases during the cycle maximum due to strong magnetic activity. Interestingly, the flow remains suppressed even after the maximum and begins to accelerate again well before the cycle minimum, reflecting the overlapping nature of solar cycles.

Simulations using this modulated flow, combined with 50\% stochastic fluctuations in the Babcock–Leighton source, reproduce realistic cycle asymmetries. Most cycles exhibit longer decay times than rise times (Figure~\ref{fig:butf}(b)). Consistent with earlier results, we find a statistically significant and strong positive correlation between the cycle amplitude and the rise rate (Figure~\ref{fig:butf}(d)), indicating that stronger cycles rise more rapidly. We also get very good and significant anticorrelation (correlation coefficient -0.57) between the cycle amplitude and the decay rate near the preceeding minima (Figure~\ref{fig:butf}(e)). However, correlations between cycle amplitude and rise time is very weak and positive.

Overall, while deterministic modulations—especially those tied to flow asymmetries and magnetic activity belts—can reproduce observed rise–decay asymmetry and rise rate correlations, Lorentz-force-driven smooth feedback alone appears insufficient.

\begin{figure*}[!ht]
\centering
\begin{tabular}{cc}
\includegraphics*[width=1.0\linewidth]{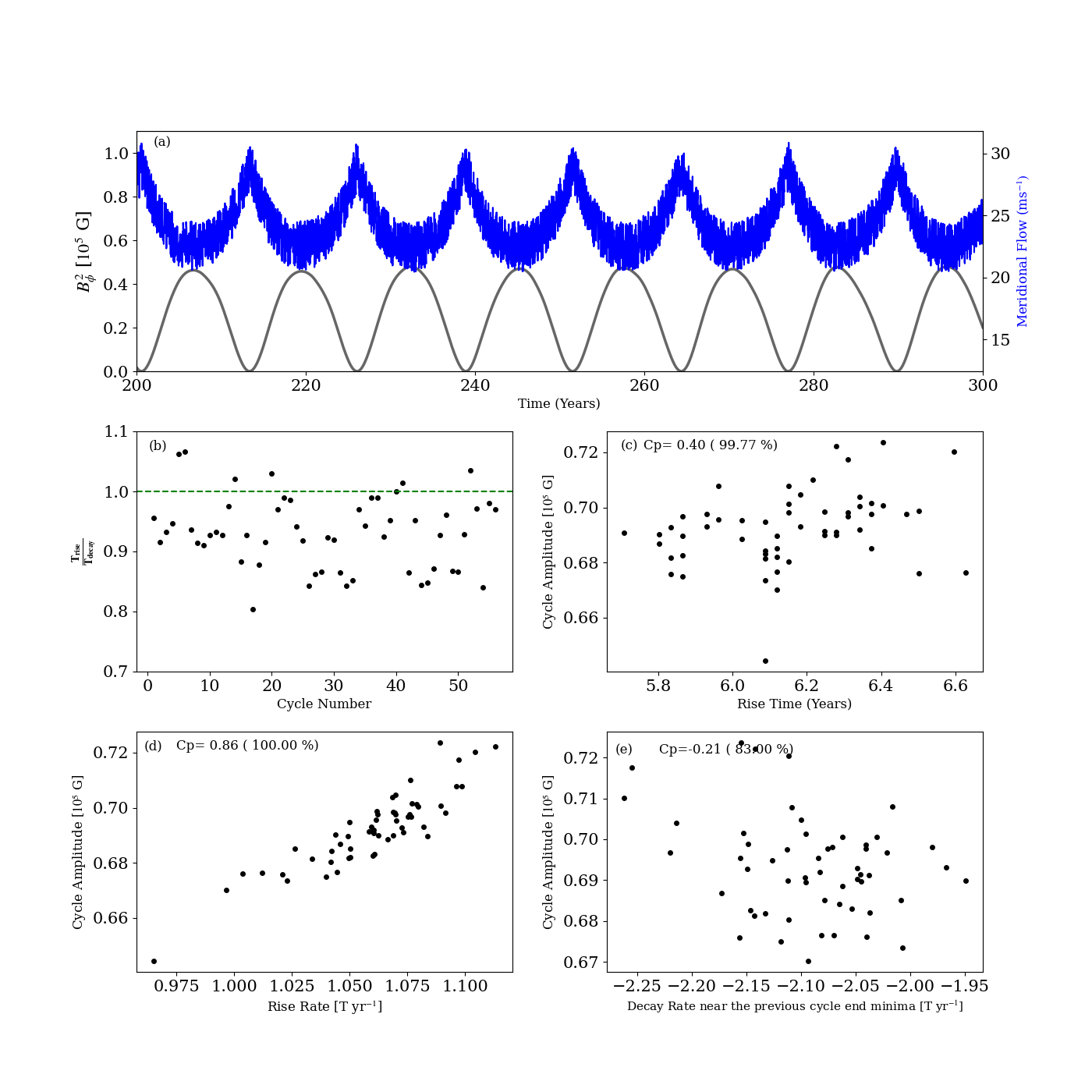}
\end{tabular}                
\caption{\footnotesize{\textbf{Figure (a), top panel:} Results from a simulation where the meridional flow is continuously modulated by Lorentz force feedback from the dynamo-generated magnetic field. A random component is superimposed on this modulation to account for convective turbulence, resulting in a hybrid meridional flow profile. The black curve (right $y$-axis) shows the evolution of magnetic energy density, while the blue curve (left $y$-axis) shows the variation in the amplitude of the meridional flow. \textbf{Figures (b)–(e):} Asymmetry measures derived from the same simulation. These four panels represent the same quantities as in Figure~\ref{fig:wald-pol}. An additional $50\%$ stochastic fluctuation has been introduced in the poloidal field generation mechanism.}}
\label{fig:lorentz-random}
\end{figure*}

\begin{figure*}[!ht]
\centering
\begin{tabular}{cc}
\includegraphics*[width=1.0\linewidth]{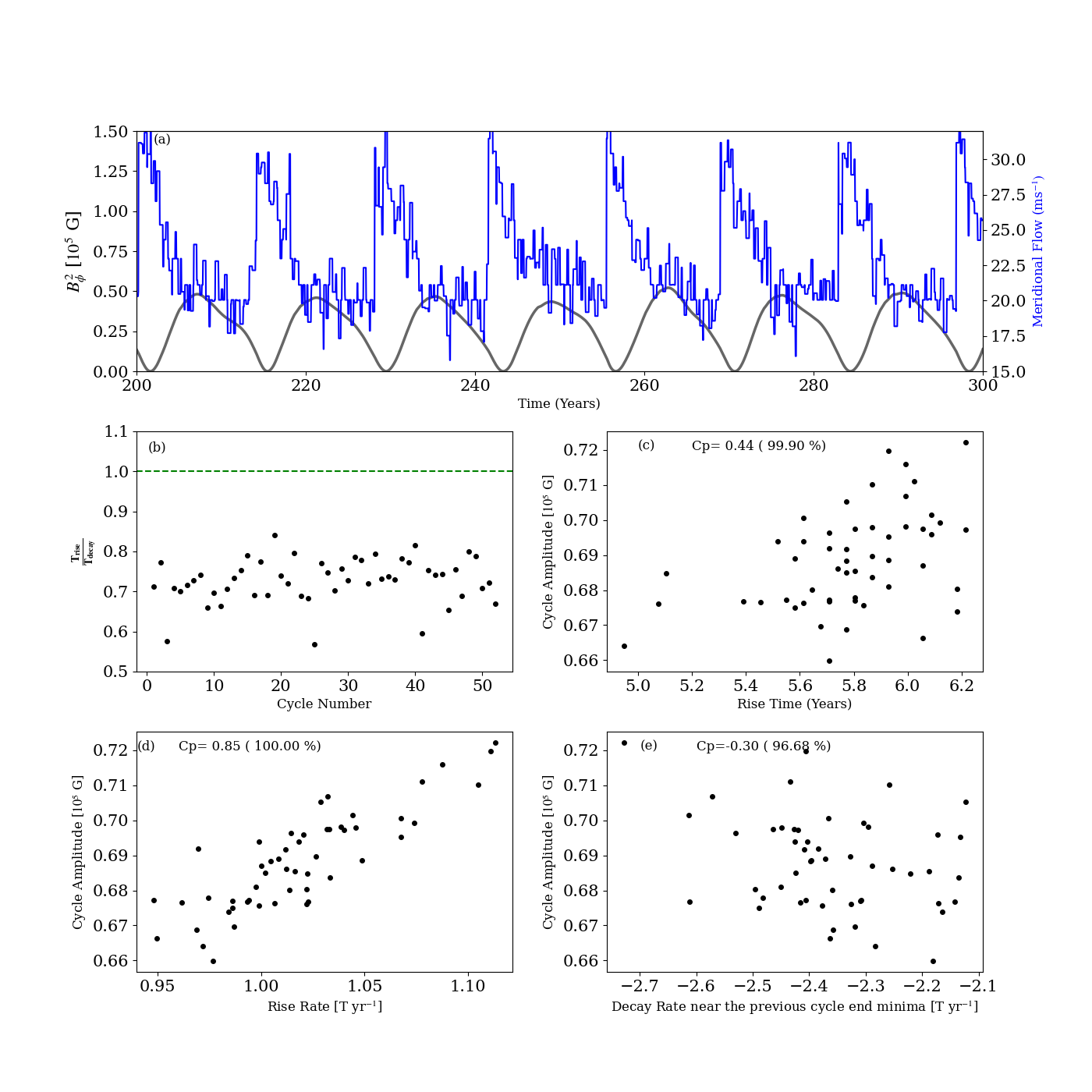}
\end{tabular}                
\caption{\footnotesize{\textbf{Figure (a), top panel:} Simulation results where the meridional flow is modulated by the active region eruption belt using the term $\delta v \sin^2(2\theta_{\mathrm{max}})$, with $\theta_{\mathrm{max}}$ denoting the latitude of peak toroidal field strength. A random component with a coherence time of three months is superimposed on this modulation to account for convective turbulence, resulting in a hybrid meridional flow profile. The black curve (right $y$-axis) shows the evolution of magnetic energy density, while the blue curve (left $y$-axis) shows the variation in the amplitude of the meridional flow.  \textbf{Figures (b)–(e):} Asymmetry measures from the same simulation, showing the same quantities as in Figure~\ref{fig:wald-pol}, with an added $50\%$ stochastic fluctuation in the poloidal field source.}}
\label{fig:eruption-random}
\end{figure*}
\subsection{Asymmetry Due to Both Deterministic and Random Variations in the Meridional Flow}
Deterministic changes in the meridional circulation—such as those induced by Lorentz-force feedback—can impose systematic temporal variations over the solar cycle, while random fluctuations, arising from turbulent convection or localized magnetic disturbances, introduce irregular and unpredictable components. In this section, we investigate how both deterministic and stochastic modulations influence the asymmetry between the rise and decay phases of solar cycles, using a combination of idealized and physically motivated flow profiles.

We first investigate the impact of purely stochastic variations by imposing a randomly fluctuating meridional circulation with a coherence time of approximately one month. This choice is motivated by helioseismic observations, which show that perturbations in the surface meridional flow evolve on timescales of weeks to months—typically spanning a few solar rotations \citep{Hathaway2010, Komm2015}. In these simulations, we also introduce stochasticity in the Babcock–Leighton source term, allowing it to fluctuate by up to 50 \%, consistent with observed scatter in sunspot tilt angles. The top panel of Figure~\ref{fig:random-merid}(a) shows the evolution of the meridional flow (blue curve) alongside the toroidal field (black curve). In this setup, rise times are often longer than decay times, and there is no consistent trend toward longer decay phases (Figure~\ref{fig:random-merid}(b)). Despite the random flow modulation, we find a strong and statistically significant positive correlation between the cycle amplitude and the rise rate, as well as a weak negative correlation between the amplitude and the decay rate near the preceding minimum (Figures~\ref{fig:random-merid}(d) and (e)). However, the correlation between rise time and cycle amplitude is weak and positive(Figure~\ref{fig:random-merid}(c)). This is consistent with previous findings by \citet{Karak2011}, who reported a significant negative amplitude–rise time correlation only when the coherence time of the meridional flow fluctuations exceeded 20 years.

In summary, while random variations in the meridional flow can impact individual cycle properties and reproduce the observed amplitude–rise rate correlation, they fail to consistently generate the typical rise–decay asymmetry. This suggests that stochastic fluctuations alone are insufficient to explain solar cycle asymmetry. 

We next explore a hybrid paradigm where both deterministic magnetic feedback and stochastic convective variability are simultaneously present. This approach acknowledges the multiscale nature of solar dynamics and offers a more comprehensive framework to describe the observed variability—including asymmetry between the rise and decay phases. Despite its plausibility, a systematic investigation of such a combined scenario has been lacking.

We begin by considering a meridional-flow profile that combines Lorentz-force–driven deterministic modulation with short-coherence stochastic variability. Concretely, we superimpose zero-mean, piecewise-constant random perturbations of amplitude $\pm 2$ $ms^{-1}$ with coherence time $\tau=2$ days onto a time-varying deterministic background set by magnetic feedback. The modulation is
\begin{equation}
v_0(t) = v_{\max}\left[1 - \tanh\!\left(\frac{|B_\phi|}{\Delta}\right)\right] + \delta v_p \,\sigma(t;\tau),
\end{equation}

where $v_{max}$ is the maximum meridional flow speed (typically near cycle minima), $\Delta$ controls controls the strength of Lorentz-force quenching near maxima, $\delta v_p$ is the velocity perturbation term and $\sigma(t, \tau)$ is the uniform random white noise between +1 and -1 with a coherence time $\tau$. The choice of fluctuation amplitude is guided by helioseismic and Doppler observations, which show that the background meridional circulation varies by a few meters per second about its mean value. A $\pm 2$ $m s^{-1}$ perturbation thus corresponds to a modest 10–20 \% fluctuation, in line with the residual variability after removing inflows around active regions. Short-coherence stochastic variability is motivated by the dynamics of near-surface convection: supergranular flows have lifetimes of order 1–2 days, and hence a coherence time of a few days is a physically reasonable proxy for unresolved turbulent forcing. We have also carried out parameter sweeps across a broader range of fluctuation amplitudes ($\delta v_p=1-5$ $ms^{-1}$) and coherence times spanning days to a month.

Figure~\ref{fig:lorentz-random} shows the results from simulations using this hybrid flow profile along with 50\% fluctuations in the Babcock–Leighton source. The top panel of Figure~\ref{fig:lorentz-random}(a) shows the resulting meridional flow (blue curve) and the corresponding toroidal field evolution (black curve). We find that in this hybrid scenario, the decay time consistently exceeds the rise time across cycles (Figure~\ref{fig:lorentz-random}(b)). We also observe a strong positive correlation between cycle amplitude and rise rate, and a statistically significant but weak negative correlation between amplitude and the decay rate near the preceding minimum (Figures~\ref{fig:lorentz-random}(d) and (e)).  However, the asymmetry remains sensitive to parameter choices: when the mean $\alpha_{BL}$ coefficient is increased or decreased, many cycles exhibit longer rise times than decay times. These results suggest that hybrid scenarios, combining magnetic feedback and convective randomness, can in some cases reproduce the observed rise–decay asymmetry. However, since the outcome depends strongly on parameter values, such approaches should be regarded as plausible but not uniquely sufficient explanations of solar cycle asymmetry.

We further develop an alternative hybrid flow profile by superimposing a random meridional flow component of amplitude $\pm 2$ m/s and one-month coherence time onto the deterministic latitude-dependent flow profile proportional to $\delta v \sin^2(2\theta_{\mathrm{max}})$, where $\theta_{\mathrm{max}}$ denotes the latitude of maximum toroidal field strength. Modulation is mathematically expressed as:
\begin{equation}
    v_{0} (t) = v_{const} + \delta v \, \sin^{2}(2\theta_{\mathrm{max}}) + \delta v_p \,\sigma(t;\tau),
\end{equation}
where, $\delta v_p$ is the velocity perturbation term and $\sigma(t;\tau)$ is the uniform white random noise between +1 and -1. Figure~\ref{fig:eruption-random} shows the simulation results for this setup, again with 50\% fluctuations in the Babcock–Leighton source term. The blue curve in the top panel of Figure~\ref{fig:eruption-random}(a) shows the resulting flow profile. Notably, this hybrid flow structure closely resembles that obtained when combining Lorentz-force feedback with short-timescale stochasticity. The flow amplitude decreases with increasing cycle strength and reaches its minimum slightly after cycle maximum—consistent with the analytic model reported by \citet{GHazra2017}.

As shown in Figure~\ref{fig:eruption-random}(b), this configuration also produces longer decay times compared to rise times, with the average rise-to-decay time ratio dropping to around 0.75. Moreover, we find a strong positive correlation between cycle amplitude and rise rate, and a very weak negative correlation between amplitude and the decay rate near the preceding minimum (Figures~\ref{fig:eruption-random}(c)–(e)).

 In summary, our results indicate that a hybrid approach—incorporating both deterministic magnetic feedback and stochastic variability in the meridional flow—can reproduce several observed features of solar cycle asymmetry. 

\section{Discussion and Conclusions}
In this study, we have investigated the role of time-dependent and spatially modulated meridional circulation in shaping the asymmetry of solar cycles using a series of kinematic flux transport dynamo simulations. Given the current lack of precise observational constraints on the structure and temporal evolution of the Sun’s internal meridional flow—particularly in the deep convection zone—we explore a range of plausible dynamical scenarios. Specifically, we have systematically investigated the role of both deterministic and stochastic variations in the meridional flow in producing the observed asymmetry between the rise and decay phases of solar cycles. Our approach involved a series of kinematic dynamo simulations incorporating various idealized and physically motivated flow modulation scenarios, in combination with stochastic fluctuations in the Babcock–Leighton poloidal source. Stochastic fluctuations in the Babcock-Leighton mechnaism reflects the inherent randomness associated with active region emergence and tilt-angle scatter. Please note that our model operates in the diffusivity dominated regime. In this regime, the cycle period depends somewhat less strongly on the meridional circulation speed, following the relation $T \propto  {\bf v_0}^{-0.75}$, compared to the advection-dominated regime, where $T \propto {\bf v_0}^{-0.89}$ \citep{Yeates2008}. The key distinction between these two transport regimes lies in the role of diffusion: in the advection-dominated case, the field is primarily transported by circulation with minimal decay, whereas in the diffusivity-dominated regime, diffusion contribute significantly in the transport mechanism and further leads to a reduction in the field strength due to diffusive decay during the transport process.

To characterize the resulting cycle asymmetry, we evaluated four key diagnostics: (1) the ratio of rise to decay times, (2) the correlation between cycle amplitude and rise time, (3) the correlation between cycle amplitude and rise rate, and (4) the correlation between cycle amplitude and the decay rate near the preceding cycle minimum. Our results show that the rise–decay asymmetry is highly sensitive to the temporal structure of the meridional flow. In deterministic scenarios with a deliberately chosen flow profile—faster during the rising phase and slower during the declining phase—the model reproduces a pronounced asymmetry, with shorter rise times and longer decay times, qualitatively matching solar observations. 

When both the meridional flow and the Babcock–Leighton mechanism are treated as stochastic, the model fails to consistently produce cycles with decay times exceeding rise times. Fluctuations in the Babcock–Leighton mechanism alone are also insufficient to reproduce the observed rise–decay asymmetry. In contrast, physically motivated, well-chosen deterministic variations of the meridional flow—motivated by Lorentz-force feedback and interpreted as a response to the emergence and equatorward migration of active regions, hence closely linked to the butterfly diagram—successfully reproduce the observed asymmetry when combined with stochastic Babcock–Leighton fluctuations. For instance, when the meridional flow varies as $\delta v \sin^2(2\theta_{\mathrm{max}})$, where $\delta v$ is the modulation amplitude and $\theta_{\mathrm{max}}$ is the latitude of the toroidal field maximum, the flow is weaker near the cycle maximum, remains weak for some time after the maximum, and then increases again, capturing the essential effect of Lorentz-force feedback. Please note that, by contrast to the dynamo model used in our study, \cite{Zhang2022} recently developed a solar dynamo model in which the meridional circulation plays a very minor role in controlling cycle periodicity and amplitude, primarily because the toroidal field generation occurs throughout the convection zone \citep{Jiang2025}. In such a model configuration, our proposed time dependent activity belt localized meridional circulation will play a lesser role in explaining the cycle asymmetry. In summary, different classes of solar dynamo models may possess different behavior depending on the combination of physical ingredients used to drive the dynamo (e.g. alpha-effect, Babcock-Leighton, meridional circulation, turbulent pumping) that can result in different degree of asymmetry they can reproduce (see for instance \cite{Jiang2007, Karak2017, Biswas2022, Jiang2020, Zhang2022}).

Hybrid scenarios, in which the meridional flow includes both deterministic and stochastic components along with stochastic Babcock–Leighton fluctuations, also yield rise–decay asymmetry, although the degree of agreement depends on parameter choices (e.g., the value of the $\alpha$-coefficient) and the specific deterministic profile. These hybrid cases highlight the interplay between systematic magnetic feedback and convective randomness as a plausible mechanism for solar cycle asymmetry. Importantly, all this approach consistently produces a strong, statistically significant positive correlation between cycle amplitude and rise rate, confirming that stronger cycles rise more rapidly and supporting the rise rate as a reliable early indicator of cycle strength. Correlations involving the decay rate near the preceding minimum remain weak, though occasionally statistically significant.

A statistically robust positive correlation between the rise rate and cycle amplitude emerges consistently across all our simulations, irrespective of whether the underlying variations in meridional flow are deterministic, stochastic, or a combination of both. This indicates that the rise rate is a more intrinsic feature of the solar dynamo mechanism itself, rather than being highly sensitive to the specific nature of temporal variability in the flow. In flux-transport dynamo models, this correlation can be traced to the role of the $\Omega$-effect, whereby differential rotation acts efficiently on the poloidal field—to regenerate the toroidal field. When the poloidal source is strong or flow conditions are favorable, the toroidal field intensifies rapidly, leading to both a stronger and faster-rising cycle. This built-in amplification pathway naturally links the early growth rate of the cycle to its eventual amplitude.

By contrast, the rise time–amplitude correlation is comparatively weaker and less consistent. This is likely due to the ambiguity in defining the rise time in the presence of noisy cycle structures—such as double peaks, overlapping cycles, or irregular shoulders near the maximum. Rise rate, as a slope-like diagnostic, bypasses much of this ambiguity by capturing the steepness of the initial growth phase directly. Also note that negative correlation between rise time and cycle amplitude is not found in sunspot area dataset.

We also investigated the correlation between cycle amplitude and the decay rate near the preceding minimum, which is predicted to be negative in some empirical studies. In some modeled scenario, we are able to reproduce good anticorrelation between the decay rate and cycle amplitude, but not always. One possible explanation lies in the treatment of stochasticity: in our model, random fluctuations in the meridional flow and the Babcock-Leighton mechanism are introduced uniformly throughout the cycle. Observationally, however, solar cycle maxima tend to exhibit higher levels of stochasticity, driven by complex active region emergence and surface dynamics, while minima are relatively quiescent but crucial in setting the seed for the next cycle. This asymmetric nature of stochasticity implies that a more physically motivated treatment—introducing different statistical properties for maxima and minima—may be necessary to fully capture this behavior.

In conclusion, our study demonstrates that plausible variations in the meridional flow can account for several key aspects of solar cycle asymmetry, including the rise–decay time asymmetry and the strong correlation between rise rate and cycle amplitude. However, direct observational constraints on the internal meridional circulation—especially its temporal and latitudinal variations—remain lacking at present. This highlights the urgent need for improved observational diagnostics of the Sun’s internal flows. Current helioseismic techniques primarily probe the surface and near-surface layers, offering limited sensitivity to flow structures deeper in the convection zone \citep{Zhao2013, Gizon2020, Mahajan2023, Fuentes2024}. Progress in helioseismology, data assimilation, and 3D dynamo modeling will therefore be crucial for constraining solar interior dynamics and advancing our understanding of solar cycle modulation. 

\section*{Acknowledgements}
The authors thank CNES Solar Orbiter and Space Weather funds, the CNRS/Sun–Earth (AT-ST) funding program, the ESA-funded SWESNET consortium, and the ERC Whole Sun (grant No. 810238) for their financial support. We also acknowledge the use of the SIDC/SILSO sunspot database

\bibliographystyle{./aa}

\begin{thebibliography}{98}
\expandafter\ifx\csname natexlab\endcsname\relax\def\natexlab#1{#1}\fi

\bibitem[{{Babcock}(1961)}]{Babcock1961}
{Babcock}, H.~W. 1961, \apj, 133, 572

\bibitem[{{Basu} \& {Antia}(2010)}]{Basu2010}
{Basu}, S. \& {Antia}, H.~M. 2010, \apj, 717, 488

\bibitem[{{Bhowmik}(2019)}]{Bhowmik2019}
{Bhowmik}, P. 2019, \aap, 632, A117

\bibitem[{{Bhowmik} \& {Nandy}(2018)}]{Bhowmik2018}
{Bhowmik}, P. \& {Nandy}, D. 2018, Nature Communications, 9, 5209

\bibitem[{{Biswas} {et~al.}(2022){Biswas}, {Karak}, \& {Cameron}}]{Biswas2022}
{Biswas}, A., {Karak}, B.~B., \& {Cameron}, R. 2022, \prl, 129, 241102

\bibitem[{{Brun} \& {Browning}(2017)}]{Brun2017}
{Brun}, A.~S. \& {Browning}, M.~K. 2017, Living Reviews in Solar Physics, 14, 4

\bibitem[{{Brun} {et~al.}(2015){Brun}, {Garc{\'\i}a}, {Houdek}, {Nandy}, \& {Pinsonneault}}]{Brun2015}
{Brun}, A.~S., {Garc{\'\i}a}, R.~A., {Houdek}, G., {Nandy}, D., \& {Pinsonneault}, M. 2015, \ssr, 196, 303

\bibitem[{{Brun} {et~al.}(2004){Brun}, {Miesch}, \& {Toomre}}]{Brun2004}
{Brun}, A.~S., {Miesch}, M.~S., \& {Toomre}, J. 2004, \apj, 614, 1073

\bibitem[{{Brun} {et~al.}(2011){Brun}, {Miesch}, \& {Toomre}}]{Brun2011}
{Brun}, A.~S., {Miesch}, M.~S., \& {Toomre}, J. 2011, \apj, 742, 79

\bibitem[{{Brun} {et~al.}(2022){Brun}, {Strugarek}, {Noraz}, {Perri}, {Varela}, {Augustson}, {Charbonneau}, \& {Toomre}}]{Brun2022}
{Brun}, A.~S., {Strugarek}, A., {Noraz}, Q., {et~al.} 2022, \apj, 926, 21

\bibitem[{{Caligari} {et~al.}(1995){Caligari}, {Moreno-Insertis}, \& {Schussler}}]{Caligari1995}
{Caligari}, P., {Moreno-Insertis}, F., \& {Schussler}, M. 1995, \apj, 441, 886

\bibitem[{{Cameron} \& {Sch{\"u}ssler}(2007)}]{Cameron2007}
{Cameron}, R. \& {Sch{\"u}ssler}, M. 2007, \apj, 659, 801

\bibitem[{{Cameron} \& {Sch{\"u}ssler}(2008)}]{Cameron2008}
{Cameron}, R. \& {Sch{\"u}ssler}, M. 2008, \apj, 685, 1291

\bibitem[{{Cameron} \& {Sch{\"u}ssler}(2010)}]{Cameron2010}
{Cameron}, R.~H. \& {Sch{\"u}ssler}, M. 2010, \apj, 720, 1030

\bibitem[{{Charbonneau}(2020)}]{Charbonneau2020}
{Charbonneau}, P. 2020, Living Reviews in Solar Physics, 17, 4

\bibitem[{{Chatterjee} {et~al.}(2004){Chatterjee}, {Nandy}, \& {Choudhuri}}]{Chatterjee2004}
{Chatterjee}, P., {Nandy}, D., \& {Choudhuri}, A.~R. 2004, \aap, 427, 1019

\bibitem[{{Choudhuri}(2003)}]{Choudhuri2003}
{Choudhuri}, A.~R. 2003, \solphys, 215, 31

\bibitem[{{Choudhuri} {et~al.}(1995){Choudhuri}, {Schussler}, \& {Dikpati}}]{Choudhuri1995}
{Choudhuri}, A.~R., {Schussler}, M., \& {Dikpati}, M. 1995, \aap, 303, L29

\bibitem[{{Dash} {et~al.}(2020){Dash}, {Bhowmik}, {Athira}, {Ghosh}, \& {Nandy}}]{Dash2020}
{Dash}, S., {Bhowmik}, P., {Athira}, B.~S., {Ghosh}, N., \& {Nandy}, D. 2020, \apj, 890, 37

\bibitem[{{Dash} {et~al.}(2023){Dash}, {Nandy}, \& {Usoskin}}]{Dash2023}
{Dash}, S., {Nandy}, D., \& {Usoskin}, I. 2023, \mnras, 525, 4801

\bibitem[{{Dikpati} \& {Charbonneau}(1999)}]{Dikpati1999}
{Dikpati}, M. \& {Charbonneau}, P. 1999, \apj, 518, 508

\bibitem[{{Dikpati} {et~al.}(2008){Dikpati}, {Gilman}, \& {de Toma}}]{Dikpati2008}
{Dikpati}, M., {Gilman}, P.~A., \& {de Toma}, G. 2008, \apjl, 673, L99

\bibitem[{{Fuentes} {et~al.}(2024){Fuentes}, {Hindman}, {Zhao}, {Blume}, {Camisassa}, {Featherstone}, {Hartlep}, {Korre}, \& {Matilsky}}]{Fuentes2024}
{Fuentes}, J.~R., {Hindman}, B.~W., {Zhao}, J., {et~al.} 2024, \apj, 961, 78

\bibitem[{{Garg} {et~al.}(2019){Garg}, {Karak}, {Egeland}, {Soon}, \& {Baliunas}}]{Garg2019}
{Garg}, S., {Karak}, B.~B., {Egeland}, R., {Soon}, W., \& {Baliunas}, S. 2019, \apj, 886, 132

\bibitem[{{Ghosh} {et~al.}(2024){Ghosh}, {Kumar}, {Prasad}, \& {Karak}}]{Ghosh2024}
{Ghosh}, A., {Kumar}, P., {Prasad}, A., \& {Karak}, B.~B. 2024, \aj, 167, 209

\bibitem[{{Ghosh} {et~al.}(2017){Ghosh}, {Tripathi}, {Gupta}, {Polito}, {Mason}, \& {Solanki}}]{Ghosh2017}
{Ghosh}, A., {Tripathi}, D., {Gupta}, G.~R., {et~al.} 2017, \apj, 835, 244

\bibitem[{{Gizon} {et~al.}(2020){Gizon}, {Cameron}, {Pourabdian}, {Liang}, {Fournier}, {Birch}, \& {Hanson}}]{Gizon2020}
{Gizon}, L., {Cameron}, R.~H., {Pourabdian}, M., {et~al.} 2020, Science, 368, 1469

\bibitem[{Gnevyshev \& Ohl(1948)}]{Gnevyshev1948}
Gnevyshev, M. \& Ohl, A. 1948, Astron. Zh, 25, 18

\bibitem[{{Hathaway}(1996)}]{Hathaway1996}
{Hathaway}, D.~H. 1996, \apj, 460, 1027

\bibitem[{{Hathaway}(2011)}]{Hathaway2011}
{Hathaway}, D.~H. 2011, \solphys, 273, 221

\bibitem[{{Hathaway}(2015)}]{Hathaway2015}
{Hathaway}, D.~H. 2015, Living Reviews in Solar Physics, 12, 4

\bibitem[{{Hathaway} \& {Rightmire}(2010)}]{Hathaway2010}
{Hathaway}, D.~H. \& {Rightmire}, L. 2010, Science, 327, 1350

\bibitem[{{Hathaway} {et~al.}(1994){Hathaway}, {Wilson}, \& {Reichmann}}]{Hathaway1994}
{Hathaway}, D.~H., {Wilson}, R.~M., \& {Reichmann}, E.~J. 1994, \solphys, 151, 177

\bibitem[{{Hazra} \& {Choudhuri}(2017)}]{GHazra2017}
{Hazra}, G. \& {Choudhuri}, A.~R. 2017, \mnras, 472, 2728

\bibitem[{{Hazra} \& {Choudhuri}(2019)}]{GHazra2019}
{Hazra}, G. \& {Choudhuri}, A.~R. 2019, \apj, 880, 113

\bibitem[{{Hazra} {et~al.}(2023){Hazra}, {Nandy}, {Kitchatinov}, \& {Choudhuri}}]{GHazra2023}
{Hazra}, G., {Nandy}, D., {Kitchatinov}, L., \& {Choudhuri}, A.~R. 2023, \ssr, 219, 39

\bibitem[{{Hazra} {et~al.}(2020){Hazra}, {Brun}, \& {Nandy}}]{SHazra2020}
{Hazra}, S., {Brun}, A.~S., \& {Nandy}, D. 2020, \aap, 642, A51

\bibitem[{{Hazra} \& {Nandy}(2016)}]{SHazra2016}
{Hazra}, S. \& {Nandy}, D. 2016, \apj, 832, 9

\bibitem[{{Hazra} \& {Nandy}(2019)}]{SHazra2019}
{Hazra}, S. \& {Nandy}, D. 2019, \mnras, 489, 4329

\bibitem[{{Hazra} {et~al.}(2014){Hazra}, {Passos}, \& {Nandy}}]{SHazra2014}
{Hazra}, S., {Passos}, D., \& {Nandy}, D. 2014, \apj, 789, 5

\bibitem[{{Hudson}(1991)}]{Hudson1991}
{Hudson}, H.~S. 1991, \solphys, 133, 357

\bibitem[{{Hung} {et~al.}(2017){Hung}, {Brun}, {Fournier}, {Jouve}, {Talagrand}, \& {Zakari}}]{Hung2017}
{Hung}, C.~P., {Brun}, A.~S., {Fournier}, A., {et~al.} 2017, \apj, 849, 160

\bibitem[{{Hung} {et~al.}(2015){Hung}, {Jouve}, {Brun}, {Fournier}, \& {Talagrand}}]{Hung2015}
{Hung}, C.~P., {Jouve}, L., {Brun}, A.~S., {Fournier}, A., \& {Talagrand}, O. 2015, \apj, 814, 151

\bibitem[{{Jaswal} {et~al.}(2024){Jaswal}, {Saha}, \& {Nandy}}]{Jaswal2024}
{Jaswal}, P., {Saha}, C., \& {Nandy}, D. 2024, \mnras, 528, L27

\bibitem[{{Jiang}(2020)}]{Jiang2020}
{Jiang}, J. 2020, \apj, 900, 19

\bibitem[{{Jiang} {et~al.}(2014){Jiang}, {Cameron}, \& {Sch{\"u}ssler}}]{Jiang2014}
{Jiang}, J., {Cameron}, R.~H., \& {Sch{\"u}ssler}, M. 2014, \apj, 791, 5

\bibitem[{{Jiang} {et~al.}(2007){Jiang}, {Chatterjee}, \& {Choudhuri}}]{Jiang2007}
{Jiang}, J., {Chatterjee}, P., \& {Choudhuri}, A.~R. 2007, \mnras, 381, 1527

\bibitem[{{Jiang} {et~al.}(2018){Jiang}, {Wang}, {Jiao}, \& {Cao}}]{Jiang2018}
{Jiang}, J., {Wang}, J.-X., {Jiao}, Q.-R., \& {Cao}, J.-B. 2018, \apj, 863, 159

\bibitem[{{Jiang} \& {Zhang}(2025)}]{Jiang2025}
{Jiang}, J. \& {Zhang}, Z. 2025, \aap, 700, A210

\bibitem[{{Jouve} \& {Brun}(2007)}]{Jouve2007}
{Jouve}, L. \& {Brun}, A.~S. 2007, \aap, 474, 239

\bibitem[{{Jouve} {et~al.}(2008){Jouve}, {Brun}, {Arlt}, {Brandenburg}, {Dikpati}, {Bonanno}, {K{\"a}pyl{\"a}}, {Moss}, {Rempel}, {Gilman}, {Korpi}, \& {Kosovichev}}]{Jouve2008}
{Jouve}, L., {Brun}, A.~S., {Arlt}, R., {et~al.} 2008, \aap, 483, 949

\bibitem[{{Jouve} {et~al.}(2011){Jouve}, {Brun}, \& {Talagrand}}]{Jouve2011}
{Jouve}, L., {Brun}, A.~S., \& {Talagrand}, O. 2011, \apj, 735, 31

\bibitem[{{Jouve} {et~al.}(2025){Jouve}, {Hung}, {Brun}, {Hazra}, {Fournier}, {Talagrand}, {Perri}, \& {Strugarek}}]{Jouve2025}
{Jouve}, L., {Hung}, C.~P., {Brun}, A.~S., {et~al.} 2025, \aap, 699, A30

\bibitem[{{Karak} \& {Choudhuri}(2011)}]{Karak2011}
{Karak}, B.~B. \& {Choudhuri}, A.~R. 2011, \mnras, 410, 1503

\bibitem[{{Karak} \& {Miesch}(2017)}]{Karak2017}
{Karak}, B.~B. \& {Miesch}, M. 2017, \apj, 847, 69

\bibitem[{{Komitov} \& {Bonev}(2001)}]{Komitov2001}
{Komitov}, B. \& {Bonev}, B. 2001, \apjl, 554, L119

\bibitem[{{Komm} {et~al.}(2015){Komm}, {Gonz{\'a}lez Hern{\'a}ndez}, {Howe}, \& {Hill}}]{Komm2015}
{Komm}, R., {Gonz{\'a}lez Hern{\'a}ndez}, I., {Howe}, R., \& {Hill}, F. 2015, \solphys, 290, 3113

\bibitem[{{Komm} {et~al.}(1993){Komm}, {Howard}, \& {Harvey}}]{Komm93}
{Komm}, R.~W., {Howard}, R.~F., \& {Harvey}, J.~W. 1993, \solphys, 147, 207

\bibitem[{{Kumar} {et~al.}(2022){Kumar}, {Biswas}, \& {Karak}}]{Kumar2022}
{Kumar}, P., {Biswas}, A., \& {Karak}, B.~B. 2022, \mnras, 513, L112

\bibitem[{{Leighton}(1969)}]{Leighton1969}
{Leighton}, R.~B. 1969, \apj, 156, 1

\bibitem[{{Lekshmi} {et~al.}(2019){Lekshmi}, {Nandy}, \& {Antia}}]{Lekshmi2019}
{Lekshmi}, B., {Nandy}, D., \& {Antia}, H.~M. 2019, \mnras, 489, 714

\bibitem[{{Lopes} \& {Passos}(2009)}]{Lopes2009}
{Lopes}, I. \& {Passos}, D. 2009, \solphys, 257, 1

\bibitem[{{Mahajan} {et~al.}(2021){Mahajan}, {Hathaway}, {Mu{\~n}oz-Jaramillo}, \& {Martens}}]{Mahajan2021}
{Mahajan}, S.~S., {Hathaway}, D.~H., {Mu{\~n}oz-Jaramillo}, A., \& {Martens}, P.~C. 2021, \apj, 917, 100

\bibitem[{{Mahajan} {et~al.}(2023){Mahajan}, {Sun}, \& {Zhao}}]{Mahajan2023}
{Mahajan}, S.~S., {Sun}, X., \& {Zhao}, J. 2023, \apj, 950, 63

\bibitem[{{Mandal} {et~al.}(2020){Mandal}, {Krivova}, {Solanki}, {Sinha}, \& {Banerjee}}]{Mandal2020}
{Mandal}, S., {Krivova}, N.~A., {Solanki}, S.~K., {Sinha}, N., \& {Banerjee}, D. 2020, \aap, 640, A78

\bibitem[{{Mazumder} {et~al.}(2018){Mazumder}, {Bhowmik}, \& {Nandy}}]{Mazumder2018}
{Mazumder}, R., {Bhowmik}, P., \& {Nandy}, D. 2018, \apj, 868, 52

\bibitem[{{Miesch} {et~al.}(2008){Miesch}, {Brun}, {DeRosa}, \& {Toomre}}]{Miesch2008}
{Miesch}, M.~S., {Brun}, A.~S., {DeRosa}, M.~L., \& {Toomre}, J. 2008, \apj, 673, 557

\bibitem[{{Nandy} {et~al.}(2023){Nandy}, {Baruah}, {Bhowmik}, {Dash}, {Gupta}, {Hazra}, {Lekshmi}, {Pal}, {Pal}, {Roy}, {Saha}, \& {Sinha}}]{Nandy2023}
{Nandy}, D., {Baruah}, Y., {Bhowmik}, P., {et~al.} 2023, Journal of Atmospheric and Solar-Terrestrial Physics, 248, 106081

\bibitem[{{Nandy} \& {Choudhuri}(2002)}]{Nandy2002}
{Nandy}, D. \& {Choudhuri}, A.~R. 2002, Science, 296, 1671

\bibitem[{{Nandy} {et~al.}(2011){Nandy}, {Mu{\~n}oz-Jaramillo}, \& {Martens}}]{Nandy2011}
{Nandy}, D., {Mu{\~n}oz-Jaramillo}, A., \& {Martens}, P. C.~H. 2011, \nat, 471, 80

\bibitem[{{Noraz} {et~al.}(2025){Noraz}, {Brun}, \& {Strugarek}}]{Noraz2025}
{Noraz}, Q., {Brun}, A.~S., \& {Strugarek}, A. 2025, \apj, 981, 206

\bibitem[{{Noraz} {et~al.}(2022){Noraz}, {Brun}, {Strugarek}, \& {Depambour}}]{Noraz2022}
{Noraz}, Q., {Brun}, A.~S., {Strugarek}, A., \& {Depambour}, G. 2022, \aap, 658, A144

\bibitem[{{Pal} {et~al.}(2023){Pal}, {Bhowmik}, {Mahajan}, \& {Nandy}}]{Pal2023}
{Pal}, S., {Bhowmik}, P., {Mahajan}, S.~S., \& {Nandy}, D. 2023, \apj, 953, 51

\bibitem[{{Pal} {et~al.}(2022){Pal}, {Nandy}, \& {Kilpua}}]{Spal2022}
{Pal}, S., {Nandy}, D., \& {Kilpua}, E. K.~J. 2022, \aap, 665, A110

\bibitem[{{Panja} {et~al.}(2021){Panja}, {Cameron}, \& {Solanki}}]{Panja2021}
{Panja}, M., {Cameron}, R.~H., \& {Solanki}, S.~K. 2021, \apj, 907, 102

\bibitem[{{Parker}(1955{\natexlab{a}})}]{Parker1955}
{Parker}, E.~N. 1955{\natexlab{a}}, \apj, 122, 293

\bibitem[{{Parker}(1955{\natexlab{b}})}]{Parker1955a}
{Parker}, E.~N. 1955{\natexlab{b}}, \apj, 121, 491

\bibitem[{{Passos} {et~al.}(2012){Passos}, {Charbonneau}, \& {Beaudoin}}]{Passos2012}
{Passos}, D., {Charbonneau}, P., \& {Beaudoin}, P. 2012, \solphys, 279, 1

\bibitem[{{Passos} {et~al.}(2014){Passos}, {Nandy}, {Hazra}, \& {Lopes}}]{Passos2014}
{Passos}, D., {Nandy}, D., {Hazra}, S., \& {Lopes}, I. 2014, \aap, 563, A18

\bibitem[{{Petrovay}(2010)}]{Petrovay2010}
{Petrovay}, K. 2010, Living Reviews in Solar Physics, 7, 6

\bibitem[{{Racine} {et~al.}(2011){Racine}, {Charbonneau}, {Ghizaru}, {Bouchat}, \& {Smolarkiewicz}}]{Racine2011}
{Racine}, {\'E}., {Charbonneau}, P., {Ghizaru}, M., {Bouchat}, A., \& {Smolarkiewicz}, P.~K. 2011, \apj, 735, 46

\bibitem[{{Rempel}(2006)}]{Rempel2006}
{Rempel}, M. 2006, \apj, 647, 662

\bibitem[{{Roy} \& {Nandy}(2023)}]{Roy2023}
{Roy}, S. \& {Nandy}, D. 2023, \apjl, 950, L11

\bibitem[{{Saha} {et~al.}(2025){Saha}, {Mukhopadhyay}, \& {Nandy}}]{Saha2025}
{Saha}, C., {Mukhopadhyay}, S., \& {Nandy}, D. 2025, \apjl, 984, L5

\bibitem[{{Samanta} {et~al.}(2019){Samanta}, {Tian}, {Yurchyshyn}, {Peter}, {Cao}, {Sterling}, {Erd{\'e}lyi}, {Ahn}, {Feng}, {Utz}, {Banerjee}, \& {Chen}}]{Samanta2019}
{Samanta}, T., {Tian}, H., {Yurchyshyn}, V., {et~al.} 2019, Science, 366, 890

\bibitem[{{Sanchez} {et~al.}(2014){Sanchez}, {Fournier}, \& {Aubert}}]{Sanchez2014}
{Sanchez}, S., {Fournier}, A., \& {Aubert}, J. 2014, \apj, 781, 8

\bibitem[{{Sarkar} {et~al.}(2017){Sarkar}, {Vaidya}, {Hazra}, \& {Bhattacharyya}}]{Sarkar2017}
{Sarkar}, A., {Vaidya}, B., {Hazra}, S., \& {Bhattacharyya}, J. 2017, \apj, 851, 120

\bibitem[{{Schatten}(2005)}]{Schatten2005}
{Schatten}, K. 2005, \grl, 32, L21106

\bibitem[{{Schrijver} {et~al.}(2015){Schrijver}, {Kauristie}, {Aylward}, {Denardini}, {Gibson}, {Glover}, {Gopalswamy}, {Grande}, {Hapgood}, {Heynderickx}, {Jakowski}, {Kalegaev}, {Lapenta}, {Linker}, {Liu}, {Mandrini}, {Mann}, {Nagatsuma}, {Nandy}, {Obara}, {Paul O'Brien}, {Onsager}, {Opgenoorth}, {Terkildsen}, {Valladares}, \& {Vilmer}}]{Schrijver2015}
{Schrijver}, C.~J., {Kauristie}, K., {Aylward}, A.~D., {et~al.} 2015, Advances in Space Research, 55, 2745

\bibitem[{{Strugarek} {et~al.}(2017){Strugarek}, {Beaudoin}, {Charbonneau}, {Brun}, \& {do Nascimento}}]{Strugarek2017}
{Strugarek}, A., {Beaudoin}, P., {Charbonneau}, P., {Brun}, A.~S., \& {do Nascimento}, J.~D. 2017, Science, 357, 185

\bibitem[{{Tripathi} {et~al.}(2021){Tripathi}, {Nandy}, \& {Banerjee}}]{Tripathi2021}
{Tripathi}, B., {Nandy}, D., \& {Banerjee}, S. 2021, \mnras, 506, L50

\bibitem[{{Ulrich}(2010)}]{Ulrich2010}
{Ulrich}, R.~K. 2010, \apj, 725, 658

\bibitem[{{Vashishth} \& {Karak}(2024)}]{Vashishth2024}
{Vashishth}, V. \& {Karak}, B.~B. 2024, \apj, 974, 6

\bibitem[{{{\v{S}}vanda} {et~al.}(2007){{\v{S}}vanda}, {Kosovichev}, \& {Zhao}}]{Svanda2007}
{{\v{S}}vanda}, M., {Kosovichev}, A.~G., \& {Zhao}, J. 2007, \apjl, 670, L69

\bibitem[{{Waldmeier}(1935)}]{Waldmeier1935}
{Waldmeier}, M. 1935, Astronomische Mitteilungen der Eidgen\&ouml;ssischen Sternwarte Zurich, 14, 105

\bibitem[{{Yeates} {et~al.}(2008){Yeates}, {Nandy}, \& {Mackay}}]{Yeates2008}
{Yeates}, A.~R., {Nandy}, D., \& {Mackay}, D.~H. 2008, \apj, 673, 544

\bibitem[{{Zhang} \& {Jiang}(2022)}]{Zhang2022}
{Zhang}, Z. \& {Jiang}, J. 2022, \apj, 930, 30

\bibitem[{{Zhao} {et~al.}(2013){Zhao}, {Bogart}, {Kosovichev}, {Duvall}, \& {Hartlep}}]{Zhao2013}
{Zhao}, J., {Bogart}, R.~S., {Kosovichev}, A.~G., {Duvall}, Jr., T.~L., \& {Hartlep}, T. 2013, \apjl, 774, L29

\end{thebibliography}

\end{document}